\documentclass[proof]{pasj00}

\begin{document}
\SetRunningHead{M. Ajiki et al.}{Strong low-$z$ emitters}
\Received{2005/00/00}
\Accepted{2005/00/00}

\title{Strong emission-line galaxies at low redshift 
in the field around the quasar SDSSp J104433.04$-$012502.2}

\author{Masaru     \textsc{Ajiki},\altaffilmark{1}
        Yasuhiro   \textsc{Shioya},\altaffilmark{1}
        Yoshiaki   \textsc{Taniguchi},\altaffilmark{1}
        Takashi    \textsc{Murayama},\altaffilmark{1}\\
        Tohru      \textsc{Nagao},\altaffilmark{2,3}
        Shunji S.  \textsc{Sasaki},\altaffilmark{1}
        Ryoko      \textsc{Sumiya},\altaffilmark{1}
        Taichi     \textsc{Morioka},\altaffilmark{1}\\
        Yuichiro   \textsc{Hatakeyama},\altaffilmark{1}
        Asuka      \textsc{Yokouchi},\altaffilmark{1}
	Mari I.    \textsc{Takahashi},\altaffilmark{1} and
        Osamu      \textsc{Koizumi}\altaffilmark{1}\\
}

\altaffiltext{1}{Astronomical Institute, Graduate School of Science, Tohoku University, \\
                 Aramaki, Aoba, Sendai 980-8578}
\altaffiltext{2}{INAF --- Osservatorio Astrofisico di Arcetri,\\
                 Largo Enrico Fermi 5, 50125 Firenze, Italy}
\altaffiltext{3}{National Astronomical Observatory, \\
	Mitaka, Tokyo 181-8588, Japan}


\maketitle

\begin{abstract}
We discuss observational properties of strong emission-line galaxies
 at low redshift found by our deep imaging survey
 for high-redshift Ly$\alpha$ emitters.
In our surveys, we used the narrowband filter,
 {\it NB816} ($\lambda_{\rm center}=8150$ \AA~ with FWHM $= 120$ \AA),
 and the intermediate-band filter,
 {\it IA827} ($\lambda_{\rm center}= 8270$ \AA~ with FWHM $= 340$ \AA). 
In this survey, 62 {\it NB816}-excess ($> 0.9$ mag) and
21 {\it IA827}-excess ($> 0.8$ mag) objects were found.
Among them, we found 20 {\it NB816}-excess and 4 {\it IA827}-excess
 Ly$\alpha$ emitter candidates. 
Therefore, it turns out that
 42 {\it NB816}-excess and 17 {\it IA827}-excess objects are
 strong emission-line objects at lower redshift.
Since 4 objects in the two low-$z$ samples
 are common, the total number of strong low-$z$ emitters is 55.
Applying our photometric redshift technique,
 we identify 7 H$\alpha$ emitters at $z \approx 0.24$,
 20 H$\beta$--[O{\sc iii}] ones at $z \approx 0.65$,
 and 11 [O{\sc ii}] ones at $z \approx 1.19$. 
However, we cannot determine reliable photometric redshifts
 of the remaining 17 emitters.
The distributions of their rest frame equivalent widths
 are consistently understood with recent studies of galaxy evolution from 
 $z\sim 1$ to $z\sim 0$.

\end{abstract}

\section{Introduction}

A number of deep imaging surveys for Ly$\alpha$ emitters (LAEs)
 at high redshift have been conducted in this decade,
 leading to the discovery of  forming galaxies beyond $z=5$
 (see for review, Taniguchi et al. 2003b; Spinrad 2004).
In these surveys, a narrowband filter with bandpass of
 $\sim 100$ \AA~ has been used.
One of the most popular narrowband filters is a filter with the central
 wavelength of $\approx$ 8150 \AA,
 because OH airglow emission lines are significantly weak
 in this wavelength range.
The use of this filter makes it possible to search for LAEs
 at $z \approx 5.7$ 
(Hu et al. 1999; Rhoads \& Malhotra 2001; Ajiki et al. 2002, 2003;
 Taniguchi et al. 2003a; Hu et al. 2004; Wang, Malhotra, \& Rhoads 2005;
 Ouchi et al. 2005; Westra et al. 2005).

In these surveys, strong emission-line objects such as starburst galaxies
and active galactic nuclei (AGN) at low redshift can be also found
(Stern et al. 2000; Fujita et al. 2003a; Ajiki et al. 2003, 2004; 
Hu et al. 2004); e.g., 
H$\alpha$ emitters at $z \approx 0.24$,
[O{\sc iii}]$\lambda$5007 emitters at $z \approx 0.63$, 
[O{\sc iii}]$\lambda$4959 emitters at $z \approx 0.65$,
H$\beta$ emitters at $z \approx 0.68$,
[O{\sc ii}] emitters at $z \approx 1.19$, and so on.
%
It is important to estimate how many strong emission-line objects
 at low redshift (hereafter we call them strong low-$z$ emitters)
 are present in such LAE searches and which kinds of strong low-$z$
 emitters are common. 
These strong low-$z$ emitters
 have large emission-line equivalent widths
 since a typical survey limit of the rest-frame equivalent width
for Ly$\alpha$ emitters at $z=5.7$, 20\AA~
 is corresponds to $\approx135$\AA~ for H$\alpha$ ones at $z=0.24$,
 $\approx100$\AA~ for [O{\sc iii}] ones at $z=0.65$,
and $\approx 61$\AA~ for [O{\sc ii}] ones  at $z=1.19$.

In this paper, we investigate observational properties of such 
strong emission-line objects at low redshift
found in our deep LAE survey with the narrowband filter,
 {\it NB816}, centered on 8150 \AA ~ with the passband of
 $\Delta\lambda_{\rm FWHM} = 120$ \AA,
 and the intermediate-band filter, {\it IA827},
 centered on 8270 \AA ~ with the passband of
 $\Delta\lambda_{\rm FWHM} = 340$ \AA,
 that were carried out by using  the prime-focus camera,
 Suprime-Cam on the 8.2 m Subaru Telescope (Ajiki et al. 2003, 2004).

In this paper, we adopt a flat universe with $\Omega_{\rm matter} = 0.3$,
$\Omega_{\Lambda} = 0.7$, and $H_0=$ 70 km s$^{-1}$ Mpc$^{-1}$. 
Throughout this paper, magnitudes are given in the AB system.

\section{Data}

\begin{figure}
\FigureFile(80mm,80mm){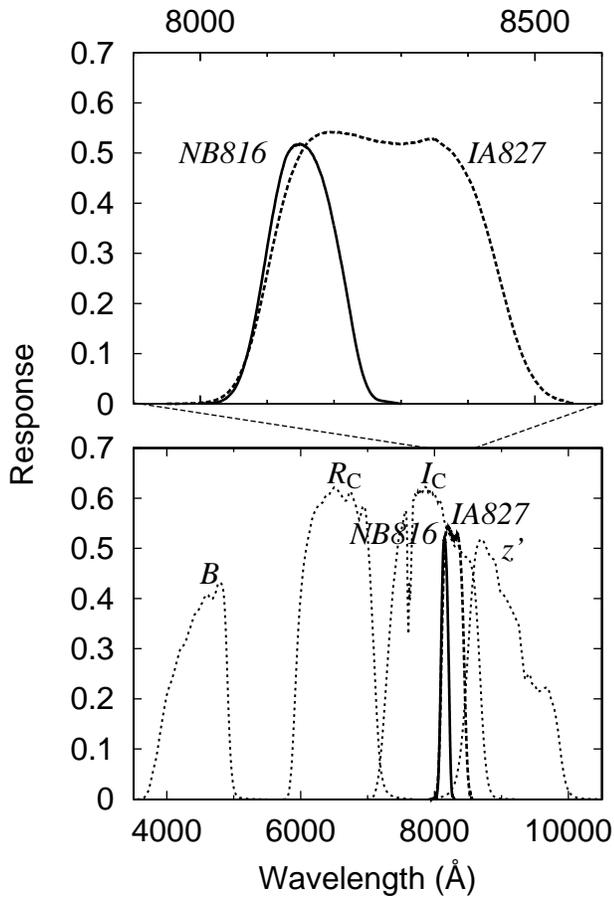}
\caption{Response curves (filter, optics, atmosphere transmission,
         and CCD sensitivity are taken into account)
         of the filters used in our observations. The upper panel shows
         the response curves of both the {\it NB816} and {\it IA827} filters.
\label{fil}}
\end{figure}

We have carried out an optical imaging survey for LAEs in the field
 surrounding the quasar SDSSp J104433.04$-$012502.2 at the redshift of 5.74
 (Fan et al. 2000; Djorgovski et al. 2001; Goodrich et al. 2001),
 using the Suprime-Cam (Miyazaki et al. 2002) on the 8.2 m
 Subaru Telescope (Kaifu et al. 2000; Iye et al. 2004) on Mauna Kea.
The Suprime-Cam consists of ten 2k $\times$ 4k CCD chips
 and provides a very wide field of view,
 $34^\prime \times 27^\prime$ (0.202 arcsec pixel$^{-1}$).
In this survey, we used the narrowband filter, {\it NB816},
 the intermediate band filter, {\it IA827},
 and the broadband filters, $B$, $R_{\rm C}$, $I_{\rm C}$, and $z^\prime$.
The limiting magnitudes (3$\sigma$) within a 2.8-arcsec aperture
 of the reduced images are {\it NB816}$=26.0$, {\it IA827}$=25.6$,
 $B=26.6$, $R_{\rm C}=26.2$, $I_{\rm C}=25.9$, and  $z^\prime=25.3$.
The total response (filter, optics, atmosphere transmission,
 and CCD sensitivity are taken into account) curve of each filter
 is shown in figure \ref{fil}.
A summary of the imaging observations, source detection, and photometry
 are given in Ajiki et al. (2003, 2004).

\section{Results and Discussion}

\subsection{{\it NB816}-selected sample of strong low-$z$ emitters}

As for {\it NB816}-excess objects, Ajiki et al. (2003) adopted
the following criteria:
\begin{eqnarray}
{\it NB816} & < & 25.0
\end{eqnarray}
and
\begin{eqnarray}
{\it Iz815} - {\it NB816} & > & 0.9,
\end{eqnarray}
where ${\it Iz815}$ is the continuum magnitude at $\lambda = 8150$ \AA~
estimated by using a combination of
the flux densities ($f_\nu$) in the $I_{\rm C}$  and $z^\prime$ bands;
\begin{eqnarray}
f_{{\it {\it Iz815}}}= 0.76 f_{I_{\rm C}} + 0.24 f_{z^\prime} \label{eq:iz1}.
\end{eqnarray}
The {\it Iz815} magnitude gives us a good approximation of the continuum
 around 8150 \AA~  for the object without significant emission or absorption lines
 in the $I_{\rm C}$ and $z^\prime$ bands.
There are 62 {\it NB816}-excess objects which satisfy the above two criteria.

In order to isolate LAE candidates from interlopers at lower
redshift, they adopted the following criteria:
\begin{eqnarray}
B & > & 26.6, \label{eq:b}\\
R_{\rm C} - I_{\rm C} & > & 1.8 \; \; \; {\rm for} \; I_{\rm C} \le 24.8,\label{eq:ri}
\end{eqnarray}
and
\begin{eqnarray}
R_{\rm C} & > & 26.6 \; \; \; {\rm for} \; I_{\rm C} > 24.8.\label{eq:r}
\end{eqnarray}
Then they obtained a final sample of 20 LAE candidates.
Two objects were identified as LAEs at $z=5.69$ and $z=5.66$ from
their follow-up optical spectroscopy
 (Ajiki et al. 2002; Taniguchi et al. 2003a).

\begin{figure}
\FigureFile(160mm,250mm){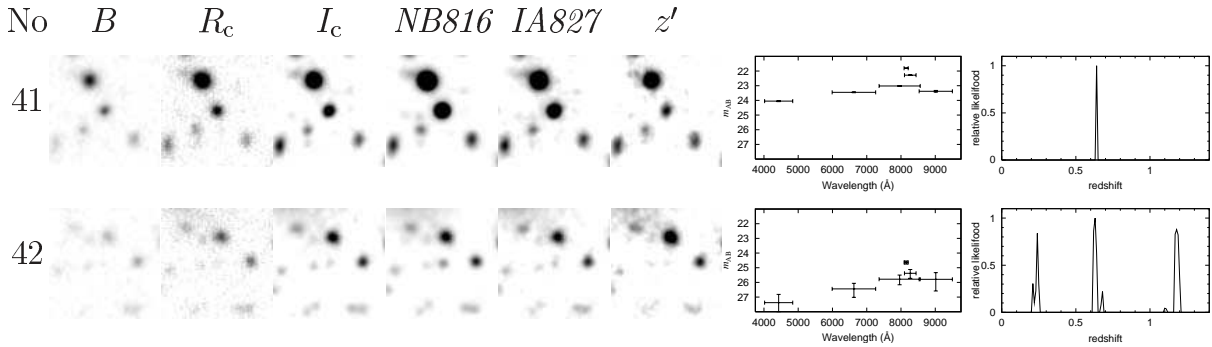}
\caption{$B$, $R_{\rm C}$, $I_{\rm C}$, {\it NB816}, {\it IA827}, and $z^\prime$ images of
         {\it NB816}-selected strong low-$z$ emitters.
         Each box is $16^{\prime \prime}$ on a side (north is up and east is left).
         The numbers shown in the left column correspond
         to those given in the first column of table \ref{tab:nb}.
         The SED and the likelihood distribution of each emitters are also shown in right two panels.
\label{thum_nb}}
\end{figure}

Their analysis indicates that the remaining 42 objects are
strong emission-line objects at lower redshift. 
Their properties are summarized in table \ref{tab:nb}.
Their images and spectral energy distributions (SEDs)
 are shown in figure \ref{thum_nb}.

\subsection{{\it IA827}-selected sample of strong low-$z$ emitters}

As for {\it IA827}-excess objects, Ajiki et al. (2004) adopted
the following criteria:
\begin{eqnarray}
{\it IA827} & < & 24.9, \\
{\it Iz827} - {\it IA827} & > & 0.8,
\end{eqnarray}
and
\begin{eqnarray}
{\it Iz827} - {\it IA827} & > & 3 \sigma({\it Iz827}-{\it IA827}),
\end{eqnarray}
where ${\it Iz827}$ is the continuum magnitude at $\lambda = 8270$ \AA~
estimated by using a combination of
the flux densities ($f_\nu$) in the $I_{\rm C}$  and $ z^\prime$ bands;
\begin{eqnarray}
f_{{\it {\it Iz827}}}= 0.64 f_{I_{\rm C}} + 0.36 f_{z^\prime}\label{eq:iz2}.
\end{eqnarray}
The {\it Iz827} magnitude gives us a good approximation of the continuum
 around 8270 \AA~  for the object without significant emission or absorption lines
 in the $I_{\rm C}$ and $z^\prime$ bands.
There are 21 {\it IA827}-excess objects which satisfy the above three criteria.
In order to isolate LAE candidates from interlopers at lower
 redshift, they adopted the criteria of (\ref{eq:b}), (\ref{eq:ri}), and (\ref{eq:r}).
Finally they obtained a final sample of 4 LAE candidates.

\begin{figure}
\FigureFile(160mm,250mm){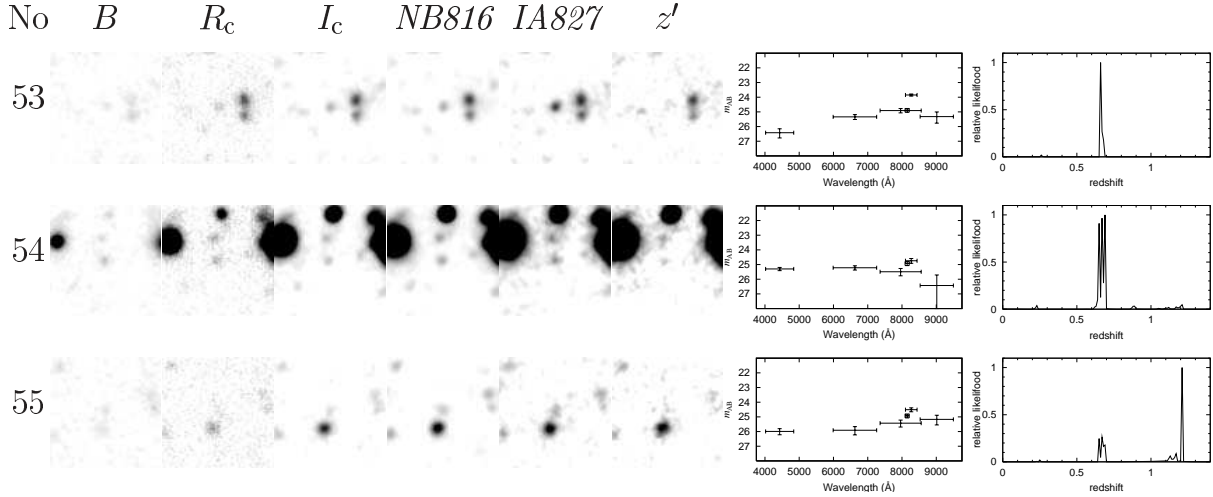}
\caption{$B$, $R_{\rm C}$, $I_{\rm C}$, {\it NB816}, {\it IA827}, and $z^\prime$ images of
         {\it IA827}-selected strong low-$z$ emitters.
         Each box is $16^{\prime \prime}$ on a side.
         The numbers shown in the left column correspond
         to those given in the first column of table \ref{tab:ia}.
         The SED and likelihood of each emitters are also shown in right two panels.
         Images of common objects, No. 2, No. 12, No. 31, and No. 41, are shown
          in figure \ref{thum_nb} (see also table \ref{tab:ia}).
\label{thum_ia}}
\end{figure}

Their analysis indicates that the remaining 17 objects are strong 
emission-line objects at lower redshift. 
Their properties are summarized in
 table \ref{tab:ia}.
Their images and SEDs are shown in figure \ref{thum_ia}.
Four of the 17 objects are common with {\it NB816}-selected emitters,
and the total number of strong low-$z$ emitters is 55.

\subsection{AGN fraction}

As well as star forming galaxies, AGNs, especially broad-line (type 1) AGNs,
 also show strong emission lines.
Therefore, we estimate a probable fraction of AGNs
 in our strong low-$z$ emitter sample from a statistical point of view.
Hao et al. (2005a) find 1317 broad-line AGNs
 from the Sloan Digital Sky Survey (SDSS; York et al. 2000).
We use the catalog of their broad-line AGN sample which is available electronically,
and find about 14\% (1.2\%), 4.5\% (0.6\%), and 1\% (0.2\%) of
 them satisfy the equivalent width criteria
 that correspond to those of our {\it NB816}- ({\it IA827}-)selected
 H$\alpha$, H$\beta$--[O{\sc iii}], and [O{\sc ii}] emitters, respectively.
Note that some of narrow line (type 2) AGNs satisfy our equivalent width criteria,
 although the fraction is negligibly small ($< 2$\%).
On the other hand, we estimate the expected number of the AGNs
 in our {\it NB816}- ({\it IA827}-) survey volume
 using AGN luminosity functions.
Hao et al. (2005b) obtained the H$\alpha$, [O{\sc iii}], and [O{\sc ii}]
 luminosity functions of the broad-line AGNs at $0<z<0.15$.
Assuming their luminosity functions,
 we find that the expected numbers of AGNs
 in our {\it NB816} ({\it IA827}) survey volume and luminosity range are
 8.0 (6.2), 2.1 (1.4), and 0.2 (0.03), for the H$\alpha$,
 H$\beta$--[O{\sc iii}], and [O{\sc ii}] emitters, respectively.
Assuming that there are no strong correlations between
 the equivalent widths and luminosities of AGNs,
 we combine these results and find that 
 the total expected number of AGNs in our sample is $\approx1$.
Since this number is only $\approx2$\% of the total number
 of our strong low-$z$ emitters,
 we ignore AGN contamination in the following analysis.

\subsection{Classification of strong low-z emitters}

Strong low-$z$ emitters are considered to be
 H$\alpha$ emitters at $z \approx 0.24$,
[O{\sc iii}] $\lambda \lambda$ 4959, 5007 emitters at $z \approx 0.63$,
 H$\beta$ emitters at $z \approx 0.68$,
 or [O{\sc ii}] $\lambda$3727 emitters at $z \approx 1.19$. 
In general, they are photometrically classified
 by using a certain broad-band color-color diagram
 unless they are not strong emitters;
 i.e., moderate emission-line objects
(e.g., Fujita et al. 2003b; Umeda et al. 2004).
However, if they are very strong emitters, 
it becomes difficult to classify them because their
strong emission lines affect their broad-band colors.
In particular, as for strong {\it NB816}-excess objects,
 we cannot use $I_{\rm C}$ band photometry in such a color analysis.

We, therefore, attempt to classify strong low-$z$ emitters
 by using the photometric redshift method 
including the effects of emission lines. 
Our basic method is given in Shioya et al.
 (2005, see also Shioya et al. 2002). 
To apply this method for strong emission-line galaxies,
 we generate SED templates with emission lines.
First, we generate
the continuum SEDs of model galaxies by GALAXEV (Bruzual \& Charlot 2003)
with adopting $\tau = 1$ Gyr.
We assume Salpeter's initial mass function (the power index of $x=1.35$)
 with the stellar mass range of $0.1 \leq m/M_{\odot} \leq 100$.
We adopt ages of $t=2$, 1, 0.1, and 0.01 Gyr and metallicity
 of $Z=0.02$, 0.008, and 0.004.
Second, we calculate the number of ionizing photons, $N_{\rm Lyc}$, 
 for each SED template.
We then evaluate H$\beta$ luminosity, $L({\rm H}\beta)$,
 using the following formula (Leitherer \& Heckman 1995): 
$L({\rm H}\beta) = 4.76 \times 10^{-13} N_{\rm Lyc} \; {\rm erg \; s^{-1}}$.
Then, we calculate emission-line ratios using Cloudy94 (Ferland 1997)
for each SED template.
We adopt emission-line ratios for the case of
 hydrogen density $n_{\rm H}=10^2 \; {\rm cm}^{-3}$,
 and ionization parameter $\log U = -2$, $-3$, and $-4$.
The gas metallicity is given the same metallicity of each SED template.
Finally we combine the calculated emission-line spectra with
 the continuum spectra for each SED template.
We adopt the dust-extinction curve for starburst galaxies
 determined by Calzetti et al. (2000) 
 with visual extinction of 0.0, 0.1, 0.3, 1.0, and 2.0.

\begin{figure}
\FigureFile(80mm,80mm){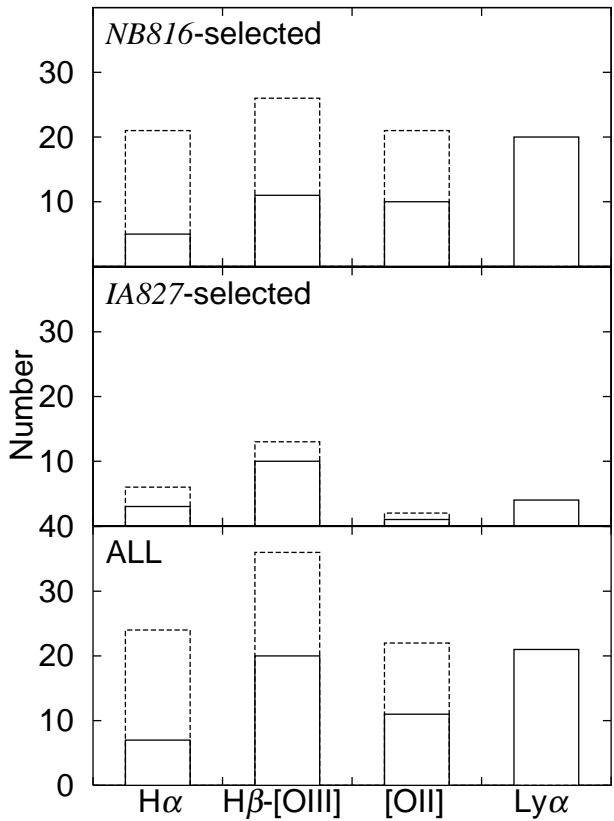}
\caption{Frequency distributions of emission-line objects.
         The solid lines show our reliable sample and dashed lines show reliable and possible sample.
\label{nhist}}
\end{figure}

\begin{figure}
\FigureFile(160mm,100mm){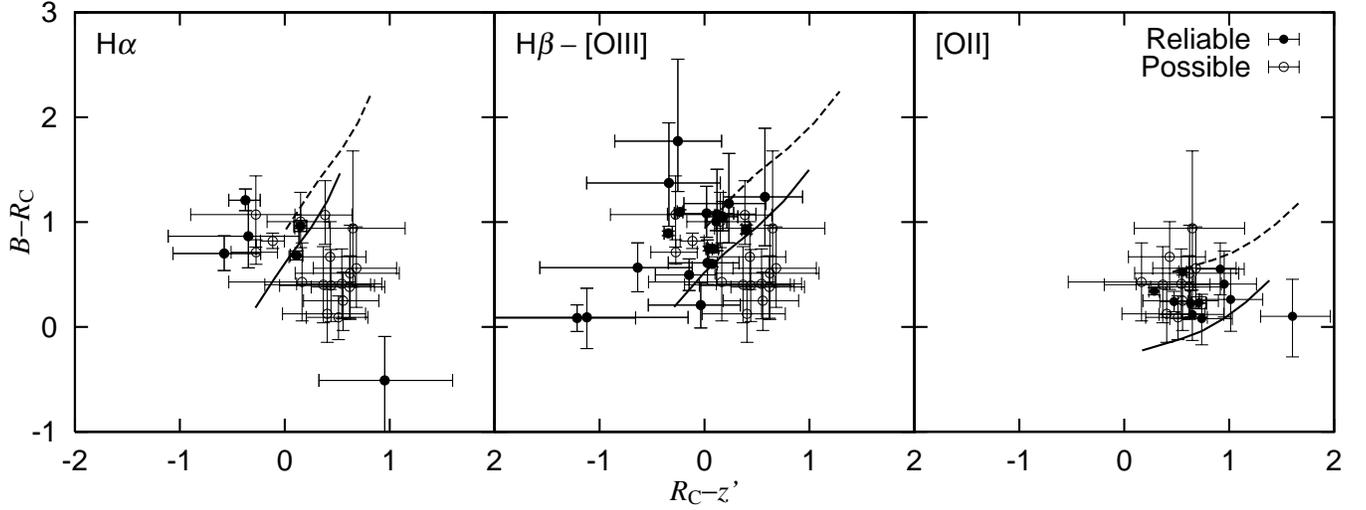}
\caption{Diagrams of our strong low-$z$ emitters
         (left: H$\alpha$; center: H$\beta$--[O{\sc iii}]; right: [O{\sc ii}])
          between $B-R_{\rm C}$ and $R_{\rm C}-z^\prime$.
         The lines are expected colors of the various age of no emission-line galaxies
         ($t=2$ -- 0.01 Gyr) 
          with and metallicity of $Z=0.02$ at $z=0.24$ (left),
          $z=0.63$ (center), and $z=1.19$ (right).
          The solid and dash lines show the case of $A_{V}=0$ and $A_{V}=1$,
          respectively.
\label{cc}}
\end{figure}

Applying this photometric redshift technique, 
we obtain the likelihood distribution for each of our 55 strong low-$z$ emitters
as a function of redshift.
The likelihood distributions are shown in right panel
 of figures \ref{thum_nb} and \ref{thum_ia}.
Note that the likelihood distributions are normalized by their maximum-peak value.
We classify each of our emitters into three emission-line types
 (H$\alpha$, H$\beta$--[O{\sc iii}], and [O{\sc ii}]).
In this classification procedure,
 we investigate likelihood distributions for each objects.
If an emitter has a peak (or peaks) higher than 0.5
 at $z\approx 0.24$, $z\approx 0.63$--0.68, or $z\approx 1.19$,
 we classify it as an H$\alpha$, an H$\beta$--[O{\sc iii}],
 or an [O{\sc ii}] emitter, respectively.
However, in some cases they are classified two or three types.
Since we cannot assign a certain emission-line type to them,
 we call them as ``possible'' sample.
On the other hand we call emitters with a certain type as ``reliable'' sample.
The photometric redshifts and types of our strong low-$z$ emitters
 are given in tables \ref{tab:nbp} and \ref{tab:iap}.
The summary of classification is also shown
 in figure \ref{nhist} and table \ref{tab:class}. 
For reference, we also show those of LAE candidates from
 Ajiki et al. (2004) in figure \ref{nhist} and table \ref{tab:class}.
In figure \ref{cc} we show color distributions of
 our strong low-$z$ emitters.
As shown in this figure, 
the colors of our strong low-$z$ emitters in each emission-line type
 are distributed around those of the model loci.
Since the distribution of observed colors of galaxies at $z=0.24$
 is very similar to that at $z=0.63$,
usual selection by color-color diagrams cannot distinguish them.
Note that possible misidentification between H$\alpha$ emitters
 and H$\beta$--[O{\sc iii}] emitters for galaxies with blue SEDs
may be unavoidable even with our photometric redshift technique
 (see section 3.6).
Follow-up spectroscopic observations are necessary for
exact classification.

\subsection{Properties of strong low-$z$ emitters found in our survey}

We estimate the line flux ($F_{\rm line}$) and observed equivalent width
 ($EW_{\rm obs}$) of each strong low-$z$ emitter
 following the method by Pascual (2001) and Fujita et al. (2003).
We add the correction of the responses of filters to their methods. 
The flux densities ($f_\nu$) in the {\it Iz815}, {\it Iz827}, {\it NB816},
 and {\it IA827} bands can be expressed as the sum of the line
flux and the continuum flux densities:
\begin{eqnarray}
f_{\it Iz815} & = & f_{\rm cont} + F_{\rm line} \frac{r_{\it Iz815}(\nu_{\rm line})}{W_{\it Iz815}},\\
f_{\it Iz827} & = & f_{\rm cont} + F_{\rm line} \frac{r_{\it Iz827}(\nu_{\rm line})}{W_{\it Iz827}},\\
f_{\it NB816} & = & f_{\rm cont} + F_{\rm line} \frac{r_{\it NB816}(\nu_{\rm line})}{W_{\it NB816}},
\end{eqnarray}
and
\begin{eqnarray}
f_{\it IA827} & = & f_{\rm cont} + F_{\rm line} \frac{r_{\it IA827}(\nu_{\rm line})}{W_{\it IA827}},
\end{eqnarray}
where $f_{\rm cont}$ is the continuum flux density
 around the emission-line wavelength.
The factors of the contributions of the line fluxes at $\nu_{\rm line}$
 to the {\it Iz815}, {\it Iz827}, {\it NB816}, and {\it IA827} filters
 are denoted as $r_{\it Iz815}(\nu_{\rm line})$,
 $r_{\it Iz827}(\nu_{\rm line})$, $r_{\it NB816}(\nu_{\rm line})$,
 and $r_{\it IA827}(\nu_{\rm line})$, respectively.
The values, $r_{\it NB816}(\nu_{\rm line})$ and $r_{\it IA827}(\nu_{\rm line})$
 are equal to the responses at $\nu_{\rm line}$
 of the {\it NB816} and {\it IA827} filters.
The values, $r_{\it Iz815}(\nu_{\rm line})$ and
 $r_{\it Iz827}(\nu_{\rm line})$, are calculated from the
 responses of the $I_{\rm C}$ and $z^\prime$ filters,
$r_{I_{\rm C}}(\nu_{\rm line})$ and $r_{z^\prime}(\nu_{\rm line})$,
 as the same manner as relation (\ref{eq:iz1}) and (\ref{eq:iz2})
as follows:
\begin{eqnarray}
r_{\it Iz815}(\nu_{\rm line}) &=& 0.76 r_{I_{\rm C}}(\nu_{\rm line}) + 0.24 r_{z^\prime}(\nu_{\rm line}),
\end{eqnarray}
and
\begin{eqnarray}
r_{\it Iz827}(\nu_{\rm line}) &=& 0.64 r_{I_{\rm C}}(\nu_{\rm line}) + 0.36 r_{z^\prime}(\nu_{\rm line}).
\end{eqnarray}
The effective widths of {\it Iz815}, {\it Iz827}, {\it NB816}, and {\it IA827}
 are denoted as $W_{\it Iz815}$, $W_{\it Iz827}$, $W_{\it NB816}$, and $W_{\it IA827}$,
respectively.
The effective width of ``{\it band}'', $W_{\it band}$, is calculated from $r_{\it band}(\nu)$
(``{\it band}'' means {\it Iz815}, {\it Iz827}, {\it NB816}, or {\it IA827}), as follows:
\begin{eqnarray}
W_{\it band} & = &  \int_{0}^{\infty} r_{\it band}(\nu) d\nu.
\end{eqnarray}
Since the ratio, $W_{\it band} / r_{\it band}(\nu)$, does not depend on the normalizing factor,
 we normalize $r_{\it band}(\nu)$ by the value averaged over effective wavelength ranges of {\it NB816}
 (8090\AA~ -- 8210\AA) for {\it Iz815} and {\it NB816}
 or those of {\it IA827} (8100\AA~ -- 8440\AA) for {\it IA827} and {\it Iz827} for convenience.
By this normalization, we obtain the effective widths, $W_{\it Iz815}=7.6 \times 10^{13}$ Hz,
 $W_{\it Iz827}=8.6 \times 10^{13}$ Hz, $W_{\it NB816}=6.5 \times 10^{12}$ Hz,
 and $W_{\it IA827}=1.7 \times 10^{13}$ Hz, respectively.
Then the continuum flux density, the line flux, and the equivalent width can be expressed as follows:
\begin{eqnarray}
f_{\rm cont}
 & = &
    \frac{  f_{\it Iz815} - f_{\it NB816}
              \frac{W_{\it NB816} / r_{\it NB816}(\nu_{\rm line})}
                   {W_{\it Iz815} / r_{\it Iz815}(\nu_{\rm line})}
         }
         {1 - 
              \frac{W_{\it NB816} / r_{\it NB816}(\nu_{\rm line})}
                   {W_{\it Iz815} / r_{\it Iz815}(\nu_{\rm line})}
         }
 \; \; {\rm for} \; {\it Iz815},
\label{eq:cnb}\\
 & = & 
    \frac{ f_{\it Iz827} -f_{\it IA827}
              \frac{W_{\it IA827} / r_{\it IA827}(\nu_{\rm line})}
                   {W_{\it Iz827} / r_{\it Iz827}(\nu_{\rm line})}
         }
         {1 - 
              \frac{W_{\it IA827} / r_{\it IA827}(\nu_{\rm line})}
                   {W_{\it Iz827} / r_{\it Iz827}(\nu_{\rm line})}
         }
 \; \; {\rm for} \; {\it Iz827},
\label{eq:cia}\\
F_{\rm line}
 & = &  (f_{\it NB816}-f_{\rm cont})
         \frac{W_{\it NB816}}{r_{\it NB816}(\nu_{\rm line})}
               \; \; {\rm for} \; {\it NB816},
 \label{eq:fnb}\\
 & = &  (f_{\it IA827}-f_{\rm cont})
        \frac{W_{\it IA827}}{r_{\it IA827}(\nu_{\rm line})}
               \; \; {\rm for} \; {\it IA827},
 \label{eq:fia}
\end{eqnarray}
and
\begin{eqnarray}
EW_{\rm obs} & = & F_{\rm line} / f_{\rm cont}\label{eq:ew}.
\end{eqnarray}
In the estimates of the continuum flux densities and line fluxes,
we use relations (\ref{eq:cnb}) and (\ref{eq:fnb}) our {\it NB816}-selected emitters
and use relations (\ref{eq:cia}) and (\ref{eq:fia}) for our {\it IA827}-selected emitters.
As for the factors of the contributions of line fluxes, 
 we use the value averaged over the effective wavelength range of {\it NB816} or {\it IA827}
[i.e, $r_{\it band}(\nu_{\rm line}) = 1$]
since we have no information about the detailed line wavelengths of our emitters.
The line fluxes, observed equivalent widths, and the FWHMs of the image sizes
 for our {\it NB816}- and {\it IA827}-selected emitters are given
 in tables \ref{tab:nbp} and \ref{tab:iap}, respectively.

The large uncertainties which is independent from photometric error
 possibly exists in the above estimates.
For example, if a very strong emission line is at nearly the edge of
 the {\it NB816} filter response curve where the response is
 $\lesssim 50$\% of the averaged value,
the equivalent width and the line flux of the emitter are underestimated by a factor of $\gtrsim 2$.
In addition to this, in the case of the H$\beta$-[O{\sc iii}] emitters,
there are also uncertainties of the types
 (H$\beta$, [O{\sc iii}]$\lambda5007$, or [O{\sc iii}]$\lambda4959$)
and the number of the emission lines covered by {\it NB816} or {\it IA827}.
For example, for a H$\beta$-[O{\sc iii}] emitter at $z=0.63$,
 we can only detect the [O{\sc iii}]$\lambda 5007$ emission line
 by our {\it NB816} or {\it IA827} observation
and miss the other (H$\beta$ and [O{\sc iii}]$\lambda 4959$)
 emission lines.
Therefore, we may underestimate $EW_{\rm obs}$ and $F_{\rm line}$
of some of our emitters by a factor of $\gtrsim 2$.
In particular, the estimates of the most of the H$\beta$--[O{\sc iii}] emitters
 may be underestimated.
The spectroscopic observation is necessary to estimate
 the equivalent widths and the line fluxes without these uncertainties.

\begin{figure}
\FigureFile(80mm,80mm){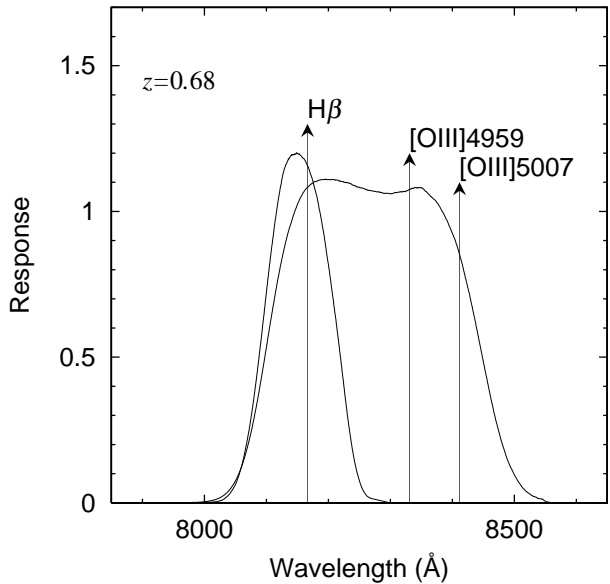}
\caption{Emission-line positions in cases of H$\beta$--[O{\sc iii}] emitters
         at $z=0.68$ compared with normalized-response curves
         of {\it NB816} and {\it IA827} filter.
\label{lineat}}
\end{figure}

We compare the estimates of $EW_{\rm obs}$ obtained
 from the {\it IA827} excess with those from the {\it NB816} excess 
 of the four emitters (Nos. 2, 12, 31, and 41) selected
 as both the {\it NB816} and {\it IA827} excess.
Although the two estimates are nearly the same for one object (No. 2),
for the other three objects (Nos. 12, 31, and 41),
the estimates from {\it IA827} excess are larger than
 those from the {\it NB816} excess by a factor of 2 -- 3.
These large differences can be explained by the uncertainty noted above.
For example, if those emitters have a emission line at $\approx 8220 $\AA
where the normalized response of {\it IA827}
 is higher than that of {\it NB816} by a factor of $\approx 2$,
 the estimates of $EW_{\rm obs}$ from {\it IA827} is also higher than
 that from {\it NB816} by a factor of $\approx 2$.
On the other hand, if they are H$\beta$--[O{\sc iii}] emitters,
 the wide passband of {\it IA827} can cover both H$\beta$ and [O{\sc iii}] emission lines
 simultaneously, while the narrower wavelength coverage of {\it NB816}
 may cover only one of them (see figure \ref{lineat}).

We also estimate the line luminosity ($L_{\rm line}$) and
 rest frame equivalent width ($EW_0$) for each strong low-$z$ emitter
 using the following relations:
\begin{eqnarray}
L_{\rm line} &=& 4 \pi d_L^2 F_{\rm line},
\end{eqnarray}
and
\begin{eqnarray}
EW_0 &=& EW_{\rm obs}/(1+z),
\end{eqnarray}
where $d_L$ is the luminosity distance.
In the above estimate, we assume $z=0.24$, $z=0.65$ and $z=1.19$
 for H$\alpha$, H$\beta$--[O{\sc iii}], and [O{\sc ii}] emitters,
 respectively.
We also estimate the absolute $B$ magnitude ($M_B$) for each of
 our H$\alpha$ and H$\beta$--[O{\sc iii}] emitters
 and the absolute magnitude at $\lambda_{\rm c}\simeq4100$\AA~($M_{410}$)
 for each of our [O{\sc ii}] emitters
 since the central wavelength of the most red-ward band used in our survey,
 $z^\prime$ ($\lambda_{\rm center}\simeq9000$\AA), corresponds to
 the rest frame wavelength of $\simeq4100$\AA~ at $z\sim1.2$.
The estimates of $L_{\rm line}$, $EW_0$, and $M_B$ (or $M_{410}$) for
 our strong low-$z$ emitters are given in tables \ref{tab:ha},
 \ref{tab:hb}, and \ref{tab:oii}.

\subsection{Comparison with emission-line galaxies at $z<0.1$ in the SDSS data}

It is interesting to study whether or not our strong low-$z$ emitters
 have common properties with those of emission-line galaxies
 in the local universe.
For this purpose, we select the three types of emitters
 in the local universe from the galaxies in the spectroscopic catalog of
 the third data release of the SDSS (SDSS DR3; Abazajian et al. 2005).
The SDSS samples are selected so as to match the absolute magnitude limits
 of our strong low-$z$ emitter samples
($M_B < -13.5$ for H$\alpha$ emitters,
 $M_B < -16.3$ for H$\beta$ -- [O{\sc iii}] emitters,
and  $M_{410} < -17.6$ for [O{\sc ii}] emitters).
The redshift ranges of the resultant SDSS samples are 
$0.003 < z < 0.006$ for H$\alpha$ emitters,
$0.008 < z < 0.015$ for H$\beta$ -- [O{\sc iii}] emitters,
and $0.03 < z < 0.06$ for [O{\sc ii}] emitters.
The rest-frame equivalent widths of H$\alpha$, H$\beta$ -- [O{\sc iii}], 
and, [O{\sc ii}] emitters in the SDSS DR3 are estimated as follows:
\begin{equation}
 EW_0=\frac{EW_{\rm H\alpha}+EW_{6548}+EW_{6583}}{1+z},
\end{equation}
\begin{equation}
 EW_0=\frac{EW_{\rm H\beta}+EW_{4959}+EW_{5007}}{1+z},
\end{equation}
and
\begin{equation}
 EW_0=\frac{EW_{3726}+EW_{3729}}{1+z},
\end{equation}
where $EW_{\rm H\alpha}$, $EW_{6548}$, $EW_{6583}$,
$EW_{\rm H\beta}$, $EW_{4959}$, $EW_{5007}$,
$EW_{3726}$, and $EW_{3729}$
 are equivalent widths of
H$\alpha$, [N{\sc ii}]$\lambda$6548, [N{\sc ii}]$\lambda$6583,
H$\beta$, [O{\sc iii}]$\lambda$4959, [O{\sc iii}]$\lambda$5007,
[O{\sc ii}]$\lambda$3726, and [O{\sc ii}]$\lambda$3729,
 respectively.
We compare the redshift range, the survey volume, the limiting $EW_0$,
 the numbers, and the number density of each type of emitters in the SDSS DR3
 to those of our strong low-$z$ emitters in table \ref{tab:survey}.
It is noted that approximately 10\% of objects are
 missed from the spectroscopic catalog of the SDSS DR3 by fiber collisions.
Since this incompleteness is negligibly small,
we perform no correction for the detection completeness the following analysis.

\begin{figure}
\FigureFile(160mm,120mm){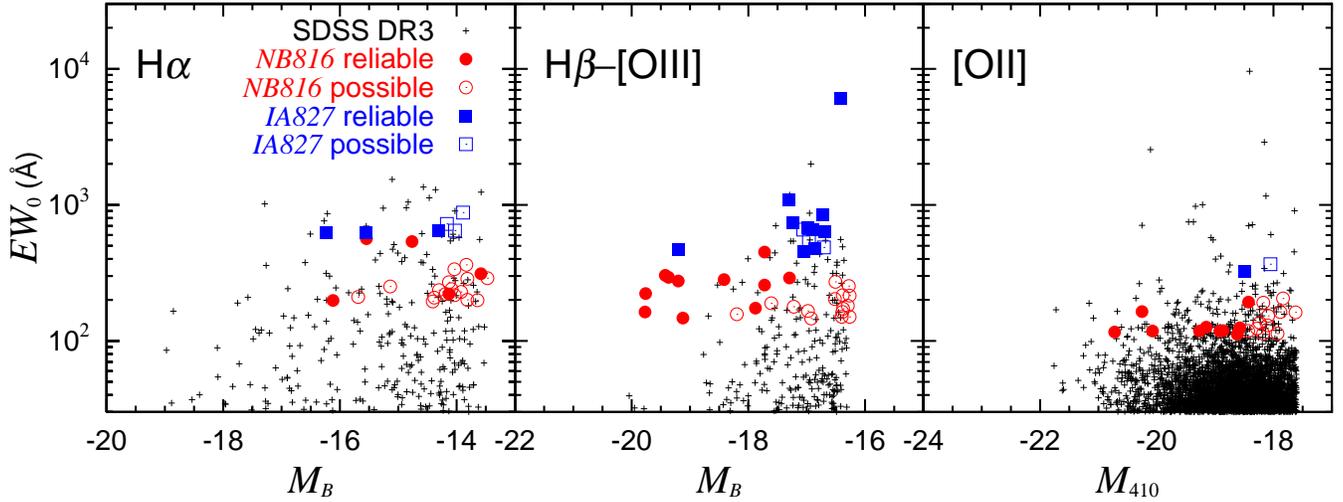}
\caption{Diagrams of $EW_0$ vs. $M_B$ or $M_{410}$ for our H$\alpha$ (left),
        H$\beta$ -- [O{\sc iii}] (center), and [O{\sc ii}] (right) emitters.
        We also show those in the SDSS DR3 catalog for the comparison.
\label{sdssEWM}}
\end{figure}

In figure \ref{sdssEWM}, we compare $EW_0$ and $M_B$ distributions for
 our sample with those of the SDSS DR3.
As shown in the left and right panels of this figure, the properties of
 $M_B$ and $EW_0$ of strong H$\alpha$ emitters at $z\approx0.24$
 and [O{\sc ii}] emitters at $z\approx1.19$ found in our survey are
 similar to those of the SDSS DR3.
However, as shown in middle panel of the figure,
 some of our H$\beta$--[O{\sc iii}] emitters at $z\approx 0.63$--0.68
 are more luminous than those of the SDSS DR3. 
We wonder if they may be misidentified in our classification,
since as mentioned before, it is often difficult distinguish between
 H$\beta$--[O{\sc iii}] and H$\alpha$ emitters.
If they were H$\alpha$ emitters at $z\sim0.24$, their $M_B$ could
 correspond to the bright part of the H$\alpha$ emitters in the SDSS DR3.
Follow-up optical spectroscopy will be necessary
to disentangle this issue.

\begin{figure}
\FigureFile(80mm,160mm){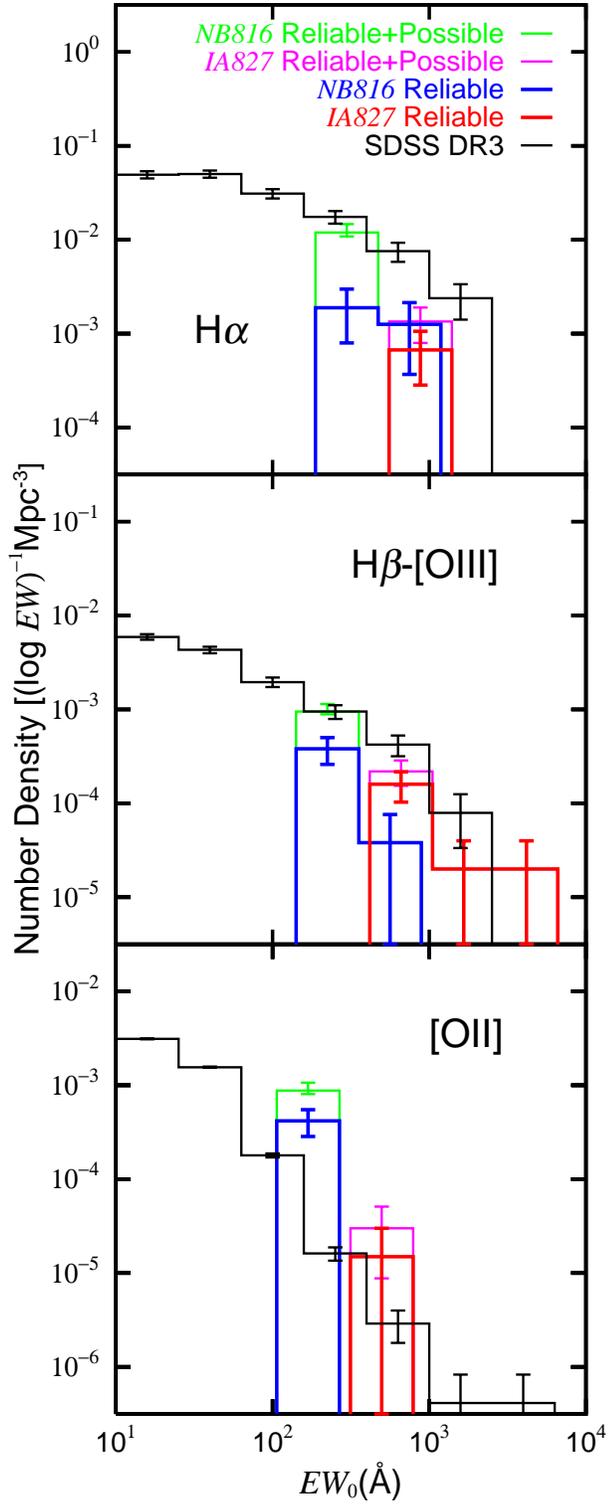}
\caption{Distributions of $EW_0$ for our H$\alpha$ (top),
         H$\beta$ -- [O{\sc iii}] (center), and [O{\sc ii}] (bottom) emitters.
        We also show those in the SDSS DR3 catalog for the comparison.
\label{sdssew}}
\end{figure}

In figure \ref{sdssew}, we show distributions of $EW_0$ of
 the three types of emitters.
Even if ``possible'' emitters included,
the number density of our H$\alpha$ emitters at $z\approx0.24$
 with $EW_0>190$\AA~($5.2 \times 10^{-3}$ Mpc$^{-3}$)
 is lower than that of the SDSS DR3 at $0.003<z<0.006$
 ($1.0 \times 10^{-2}$  Mpc$^{-3}$).
It indicates that the detection completeness of our survey
 for H$\alpha$ emitters is $\approx52$\% of the SDSS DR3.
Note that even if all of our emitters classified as H$\beta$--[O{\sc iii}]
 were to be H$\alpha$ emitters,
 the detection completeness of our survey for H$\alpha$ emitters would be
  $\approx80$\% of the SDSS DR3 at most.
On the other hand, even if excluding the ``possible'' objects,
 the number density for our [O{\sc ii}] emitters at $z\approx 1.19$ with
 $EW_0>110$\AA~($1.7 \times 10^{-4}$ Mpc$^{-3}$)
 is about 6 times higher than that for the SDSS DR3 at $0.03<z<0.06$
($3.0 \times 10^{-5}$ Mpc$^{-5}$)
 (see the bottom panel of figure \ref{sdssew} or table \ref{tab:survey}).
If we assume that the ratio of the detection completeness
 for [O{\sc ii}] emitters of our survey to that of the SDSS DR3
is almost the same as that for H$\alpha$ emitters (80\% at most),
the number density (or emission-line equivalent widths) of strong
 [O{\sc ii}] emitters may decrease from $z\sim1$ to $z\sim 0$
 by a factor of $\gtrsim7$.
Therefore, the number density of strong [O{\sc ii}] emitters shows
 strong evolution by a factor of $\gtrsim 6$--7 between $z\sim0$ and $z\sim1$.
This evolution in the [O{\sc ii}] equivalent width have been reported by
 the spectroscopic studies (e.g., Cowie et al. 1996; Hammer et al. 1997).
Although our samples have uncertainty in the detection completeness and photometric redshift,
this evolution is supported by the star formation history
between $z\sim0$ and $z\sim1$ (e.g., Tresse et al. 2002; Lilly et al. 1996).

\bigskip

We would like to thank both the Subaru staff members for
their invaluable help and T. Hayashino for his technical help.
This work was financially supported in part by
the Ministry of Education, Culture, Sports, Science and Technology
(Nos.10044052, and 10304013) and JSPS (Nos. 15340059, and 17253001). 
Data reduction/analysis was in part carried out on "sb" computer system
operated by the Astronomical Data Analysis Center (ADAC) and Subaru Telescope of
the National Astronomical Observatory of Japan.
MA, TN, and SSS are JSPS fellows.

\begin{table}
\begin{center}
\caption{Photometric properties of {\it NB816}-selected strong low-$z$ emitters. \label{tab:nb}}
\renewcommand\arraystretch{0.75}
\begin{tabular}{ccccccccc}
\hline
\hline
No. &
$\alpha$(J2000) &
$\delta$(J2000) &
$B$\footnotemark[$*$] &
$R_{\rm C}$\footnotemark[$*$] &
$I_{\rm C}$\footnotemark[$*$]  &
{\it NB816}\footnotemark[$*$]  &
{\it IA827}\footnotemark[$*$]  &
$z^\prime$\footnotemark[$*$] \\
   &
 h ~~ m ~~ s  &
$^\circ$ ~~ $^\prime$ ~~ $^{\prime\prime}$ &
&
&
&
&
&
 \\
\hline
 1 & 10 43 28.0 & -01 44 13 & 24.54 & 24.31 & 23.91 & 22.87 & 23.43 & 23.60 \\
 2\footnotemark[$\dagger$]
   & 10 43 31.4 & -01 42 36 & 25.52 & 24.32 & 24.34 & 22.88 & 23.61 & 24.69 \\
 3 & 10 43 31.8 & -01 41 39 & 26.45 & 26.05 & 25.25 & 24.22 & 24.67 & 25.09 \\
 4 & 10 43 32.1 & -01 17 44 & 23.78 & 22.89 & 22.97 & 21.68 & 22.28 & 23.23 \\
 5 & 10 43 33.0 & -01 13 46 & 26.32 & 26.06 & 25.58 & 24.50 & 25.02 & 25.05 \\
 6 & 10 43 34.8 & -01 33 19 & 26.52 & 26.11 & 25.92 & 24.82 & 25.51 & 25.56 \\
 7 & 10 43 35.0 & -01 13 47 & 25.53 & 24.82 & 24.72 & 23.74 & 24.22 & 25.09 \\
 8 & 10 43 36.5 & -01 41 04 & 24.04 & 23.70 & 23.22 & 22.12 & 22.76 & 23.42 \\
 9 & 10 43 36.9 & -01 15 18 & 26.39 & 26.00 & 25.95 & 24.81 & 24.87 & 25.39 \\
10 & 10 43 39.7 & -01 13 39 & 23.66 & 22.57 & 22.63 & 21.51 & 22.04 & 22.80 \\
11 & 10 43 42.0 & -01 14 07 & 25.90 & 25.20 & 25.35 & 23.93 & 24.66 & 25.77 \\
12\footnotemark[$\dagger$]
   & 10 43 43.0 & -01 35 52 & 26.60 & 26.20 & 25.69 & 24.45 & 24.79 & 25.83 \\
13 & 10 43 43.5 & -01 43 02 & 26.11 & 25.99 & 25.77 & 24.62 & 25.45 & 25.58 \\
14 & 10 43 44.0 & -01 26 06 & 26.60 & 25.93 & 26.19 & 24.75 & 25.43 & 25.50 \\
15 & 10 43 45.5 & -01 43 13 & 24.03 & 23.50 & 22.69 & 21.80 & 22.25 & 22.95 \\
16 & 10 43 47.0 & -01 21 57 & 26.45 & 26.35 & 25.09 & 24.04 & 24.48 & 24.75 \\
17 & 10 43 50.3 & -01 27 59 & 26.65 & 25.58 & 24.94 & 23.97 & 24.65 & 25.20 \\
18 & 10 43 50.4 & -01 44 35 & 26.87 & 26.36 & 25.97 & 24.98 & 25.51 & 25.74 \\
19 & 10 43 57.5 & -01 44 19 & 26.01 & 25.89 & 25.65 & 24.30 & 24.88 & 25.24 \\
20 & 10 44 00.5 & -01 15 24 & 25.38 & 25.15 & 24.95 & 23.83 & 24.18 & 24.52 \\
21 & 10 44 02.2 & -01 30 30 & 26.14 & 25.52 & 24.87 & 23.68 & 24.45 & 25.50 \\
22 & 10 44 02.8 & -01 23 49 & 25.82 & 24.77 & 24.35 & 23.38 & 23.83 & 24.59 \\
23 & 10 44 07.2 & -01 21 38 & 26.49 & 25.49 & 25.41 & 24.47 & 24.80 & 25.34 \\
24 & 10 44 09.4 & -01 29 38 & 26.28 & 26.79 & 25.32 & 24.23 & 25.14 & 25.84 \\
25 & 10 44 11.5 & -01 15 06 & 25.72 & 25.22 & 24.40 & 23.33 & 24.07 & 25.37 \\
26 & 10 44 12.7 & -01 42 36 & 26.74 & 26.18 & 25.85 & 24.82 & 25.27 & 25.49 \\
27 & 10 44 17.1 & -01 18 05 & 26.27 & 25.72 & 25.00 & 23.99 & 24.41 & 24.80 \\
28 & 10 44 18.1 & -01 27 59 & 23.69 & 22.94 & 22.43 & 21.54 & 22.14 & 22.90 \\
29 & 10 44 22.2 & -01 19 12 & 26.64 & 26.21 & 26.04 & 24.90 & 25.49 & 26.05 \\
30 & 10 44 36.2 & -01 23 44 & 24.27 & 23.53 & 23.10 & 22.25 & 22.62 & 23.45 \\
31\footnotemark[$\dagger$]
   & 10 44 36.5 & -01 18 51 & 26.77 & 25.70 & 25.23 & 24.38 & 24.55 & 25.97 \\
32 & 10 44 44.5 & -01 34 40 & 26.66 & 25.79 & 25.07 & 24.26 & 24.83 & 26.14 \\
33 & 10 44 46.8 & -01 26 19 & 25.84 & 25.75 & 25.32 & 24.34 & 24.92 & 25.24 \\
34 & 10 44 48.0 & -01 38 27 & 26.22 & 25.65 & 24.31 & 22.98 & 23.91 & 26.29 \\
35 & 10 44 51.6 & -01 35 04 & 26.24 & 25.99 & 26.11 & 24.87 & 25.35 & 25.43 \\
36 & 10 44 52.4 & -01 13 12 & 25.34 & 24.33 & 23.77 & 22.56 & 23.61 & 24.22 \\
37 & 10 44 52.7 & -01 37 22 & 24.75 & 23.79 & 23.56 & 22.65 & 23.13 & 23.64 \\
38 & 10 44 54.6 & -01 20 15 & 25.07 & 24.25 & 24.12 & 23.21 & 23.68 & 24.36 \\
39 & 10 44 56.7 & -01 17 13 & 24.52 & 23.59 & 22.75 & 21.52 & 22.39 & 23.19 \\
40 & 10 44 58.3 & -01 25 31 & 25.12 & 24.88 & 24.51 & 23.53 & 23.96 & 24.40 \\
41\footnotemark[$\dagger$]
   & 10 45 00.6 & -01 38 30 & 24.05 & 23.45 & 23.02 & 21.81 & 22.28 & 23.38 \\
42 & 10 45 02.8 & -01 30 53 & 27.38 & 26.44 & 25.78 & 24.64 & 25.38 & 25.79 \\
\hline
\multicolumn{9}{@{}l@{}}{\hbox to 0pt{\parbox{130mm}{\footnotesize
\par\noindent
\footnotemark[$*$] AB magnitude in a 2$\farcs$8 diameter.
\par\noindent
\footnotemark[$\dagger$] Objects selected both {\it NB816} and {\it IA827}.
}\hss}}
\end{tabular}
\end{center}
\end{table}

\begin{table}
\begin{center}
\caption{Photometric properties of {\it IA827}-selected strong low-$z$ emitters. \label{tab:ia}}
\begin{tabular}{ccccccccc}
\hline
\hline
No. &
$\alpha$(J2000) &
$\delta$(J2000) &
$B$\footnotemark[$*$] &
$R_{\rm C}$\footnotemark[$*$] &
$I_{\rm C}$\footnotemark[$*$]  &
{\it NB816}\footnotemark[$*$]  &
{\it IA827}\footnotemark[$*$]  &
$z^\prime$\footnotemark[$*$] \\
   &
 h ~~ m ~~ s  &
$^\circ$ ~~ $^\prime$ ~~ $^{\prime\prime}$ &
&
&
&
&
&
 \\
\hline
 2\footnotemark[$\dagger$]
   & 10 43 31.4 & -01 42 36 & 25.52 & 24.32 & 24.34 & 22.88 & 23.61 & 24.69 \\
12\footnotemark[$\dagger$]
   & 10 43 43.0 & -01 35 52 & 26.60 & 26.20 & 25.69 & 24.45 & 24.79 & 25.83 \\
31\footnotemark[$\dagger$]
   & 10 44 36.5 & -01 18 51 & 26.77 & 25.70 & 25.23 & 24.38 & 24.55 & 25.97 \\
41\footnotemark[$\dagger$]
   & 10 45 00.6 & -01 38 30 & 24.05 & 23.45 & 23.02 & 21.81 & 22.28 & 23.38 \\
43 & 10 43 34.8 & -01 13 13 & 26.91 & 25.83 & 25.34 & 25.05 & 24.61 & 25.71 \\
44 & 10 43 35.5 & -01 23 10 & 26.54 & 26.15 & 25.00 & 25.52 & 24.22 & 25.71 \\
45 & 10 43 49.4 & -01 30 51 & 27.30 & 26.06 & 24.81 & 25.07 & 23.97 & 25.48 \\
46 & 10 43 51.3 & -01 17 21 & 99.00 & 25.37 & 25.42 & 25.62 & 24.73 & 26.00 \\
47 & 10 44 05.3 & -01 30 39 & 26.18 & 26.09 & 25.87 & 25.46 & 24.55 & 27.21 \\
48 & 10 44 19.4 & -01 14 01 & 27.46 & 25.69 & 25.61 & 25.15 & 24.62 & 25.94 \\
49 & 10 44 28.6 & -01 38 47 & 27.19 & 25.82 & 25.11 & 25.20 & 24.54 & 26.16 \\
50 & 10 44 43.0 & -01 36 35 & 24.42 & 23.73 & 23.21 & 22.47 & 22.49 & 23.62 \\
51 & 10 44 51.6 & -01 33 42 & 25.94 & 25.74 & 25.35 & 25.25 & 24.49 & 25.77 \\
52 & 10 44 53.5 & -01 13 52 & 27.02 & 25.84 & 25.59 & 25.27 & 24.62 & 25.61 \\
53 & 10 44 58.7 & -01 16 01 & 26.42 & 25.34 & 24.91 & 24.89 & 23.85 & 25.32 \\
54 & 10 45 08.4 & -01 14 22 & 25.30 & 25.22 & 25.49 & 24.93 & 24.74 & 26.43 \\
55 & 10 45 08.5 & -01 13 18 & 25.99 & 25.91 & 25.43 & 24.93 & 24.51 & 25.17 \\
\hline
\multicolumn{9}{@{}l@{}}{\hbox to 0pt{\parbox{130mm}{\footnotesize
\par\noindent
\footnotemark[$*$] AB magnitude in a 2$\farcs$8 diameter.
\par\noindent
\footnotemark[$\dagger$] Objects selected both {\it NB816} and {\it IA827}.
}\hss}}
\end{tabular}
\end{center}
\end{table}

\begin{table}
\begin{center}
\renewcommand\arraystretch{0.75}
\caption{Properties of {\it NB816}-selected strong low-$z$ emitters. \label{tab:nbp}}
\begin{tabular}{cccccc}
\hline
\hline
No. & Class\footnotemark[$*$] & $z_{\rm ph}$\footnotemark[$\dagger$] & $FWHM_{\rm obj}$\footnotemark[$\ddagger$] & $F_{\rm line}$\footnotemark[$\S$] & $EW_{\rm obs}$\footnotemark[$\|$] \\
    &       &  & (arcsec) & (erg s$^{-1}$ cm$^{-2}$)  & (\AA) \\
\hline
1 & [O{\sc ii}] & 1.19  & 1.2  & $1.1 \times 10^{-16}$ & $258^{+26}_{-23}$ \\
2 & H$\alpha$ & 0.24  & 1.0  & $1.4 \times 10^{-16}$ & $700^{+118}_{-93}$ \\
3 & [O{\sc ii}] & 1.20  & 1.2  & $3.2 \times 10^{-17}$ & $273^{+114}_{-77}$ \\
4 & H$\beta$--[O{\sc iii}] & 0.68  & 1.1  & $3.9 \times 10^{-16}$ & $501^{+21}_{-20}$ \\
5 & [O{\sc ii}] & 1.19  & 1.1  & $2.4 \times 10^{-17}$ & $246^{+136}_{-84}$ \\
6 & ([O{\sc ii}], H$\beta$--[O{\sc iii}], H$\alpha$) & 1.18  & 1.6  & $1.9 \times 10^{-17}$ & $284^{+256}_{-124}$ \\
7 & (H$\beta$--[O{\sc iii}], H$\alpha$) & 0.64  & 1.1  & $5.2 \times 10^{-17}$ & $311^{+80}_{-61}$ \\
8 & [O{\sc ii}] & 1.18  & 1.0  & $2.4 \times 10^{-16}$ & $360^{+20}_{-18}$ \\
9 & ([O{\sc ii}], H$\alpha$, H$\beta$--[O{\sc iii}]) & 1.20  & 1.2  & $1.8 \times 10^{-17}$ & $269^{+233}_{-117}$ \\
10 & H$\beta$--[O{\sc iii}] & 0.68  & 1.3  & $4.3 \times 10^{-16}$ & $368^{+11}_{-11}$ \\
11 & H$\alpha$ & 0.24  & 1.1  & $5.2 \times 10^{-17}$ & $666^{+361}_{-194}$ \\
12 & (H$\beta$--[O{\sc iii}], H$\alpha$, [O{\sc ii}]) & 0.64  & 1.3  & $3.0 \times 10^{-17}$ & $448^{+341}_{-164}$ \\
13 & ([O{\sc ii}], H$\alpha$, H$\beta$--[O{\sc iii}]) & 1.18  & 1.2  & $2.4 \times 10^{-17}$ & $335^{+255}_{-130}$ \\
14 & (H$\beta$--[O{\sc iii}], H$\alpha$, [O{\sc ii}]) & 0.64  & 1.1  & $2.2 \times 10^{-17}$ & $417^{+467}_{-183}$ \\
15 & [O{\sc ii}] & 1.17  & 1.0  & $2.9 \times 10^{-16}$ & $254^{+9}_{-9}$ \\
16 & [O{\sc ii}] & 1.20  & 1.3  & $3.7 \times 10^{-17}$ & $258^{+86}_{-63}$ \\
17 & (H$\beta$--[O{\sc iii}], H$\alpha$) & 0.62  & 1.0  & $4.1 \times 10^{-17}$ & $293^{+95}_{-69}$ \\
18 & ([O{\sc ii}], H$\alpha$, H$\beta$--[O{\sc iii}]) & 1.18  & 1.4  & $1.5 \times 10^{-17}$ & $247^{+260}_{-120}$ \\
19 & [O{\sc ii}] & 1.19  & 1.0  & $3.4 \times 10^{-17}$ & $423^{+250}_{-137}$ \\
20 & [O{\sc ii}] & 1.20  & 1.4  & $4.6 \times 10^{-17}$ & $276^{+76}_{-57}$ \\
21 & H$\beta$--[O{\sc iii}] & 0.63  & 1.1  & $6.2 \times 10^{-17}$ & $479^{+146}_{-101}$ \\
22 & H$\beta$--[O{\sc iii}] & 0.64  & 1.0  & $7.0 \times 10^{-17}$ & $288^{+50}_{-41}$ \\
23 & (H$\alpha$, H$\beta$--[O{\sc iii}]) & 0.24  & 1.3  & $2.4 \times 10^{-17}$ & $241^{+129}_{-81}$ \\
24 & H$\alpha$ & 0.21  & 1.0  & $3.5 \times 10^{-17}$ & $387^{+198}_{-117}$ \\
25 & H$\beta$--[O{\sc iii}] & 0.62  & 1.2  & $8.3 \times 10^{-17}$ & $423^{+83}_{-64}$ \\
26 & ([O{\sc ii}], H$\alpha$, H$\beta$--[O{\sc iii}]) & 1.19  & 1.3  & $1.8 \times 10^{-17}$ & $248^{+209}_{-108}$ \\
27 & [O{\sc ii}] & 1.20  & 1.3  & $3.9 \times 10^{-17}$ & $261^{+83}_{-61}$ \\
28 & H$\beta$--[O{\sc iii}] & 0.63  & 1.1  & $3.7 \times 10^{-16}$ & $269^{+7}_{-8}$ \\
29 & (H$\alpha$, H$\beta$--[O{\sc iii}], [O{\sc ii}]) & 0.24  & 1.1  & $1.8 \times 10^{-17}$ & $355^{+437}_{-168}$ \\
30 & H$\beta$--[O{\sc iii}] & 0.64  & 1.6  & $1.9 \times 10^{-16}$ & $243^{+13}_{-13}$ \\
31 & (H$\alpha$, H$\beta$--[O{\sc iii}]) & 0.25  & 1.7  & $2.7 \times 10^{-17}$ & $274^{+138}_{-87}$ \\
32 & H$\alpha$ & 0.23  & 1.0  & $3.1 \times 10^{-17}$ & $274^{+119}_{-80}$ \\
33 & ([O{\sc ii}], H$\alpha$) & 1.18  & 1.6  & $2.8 \times 10^{-17}$ & $258^{+120}_{-79}$ \\
34 & H$\beta$--[O{\sc iii}] & 0.62  & 1.0  & $1.3 \times 10^{-16}$ & $740^{+145}_{-110}$ \\
35 & ([O{\sc ii}], H$\alpha$, H$\beta$--[O{\sc iii}]) & 1.19  & 1.0  & $1.8 \times 10^{-17}$ & $298^{+302}_{-136}$ \\
36 & H$\beta$--[O{\sc iii}] & 0.62  & 1.1  & $1.7 \times 10^{-16}$ & $465^{+44}_{-38}$ \\
37 & H$\alpha$ & 0.24  & 1.0  & $1.3 \times 10^{-16}$ & $245^{+20}_{-18}$ \\
38 & (H$\alpha$, H$\beta$--[O{\sc iii}]) & 0.24  & 1.0  & $7.9 \times 10^{-17}$ & $260^{+37}_{-32}$ \\
39 & H$\beta$--[O{\sc iii}] & 0.62  & 1.3  & $4.5 \times 10^{-16}$ & $484^{+17}_{-16}$ \\
40 & [O{\sc ii}] & 1.20  & 1.1  & $5.9 \times 10^{-17}$ & $258^{+50}_{-41}$ \\
41 & H$\beta$--[O{\sc iii}] & 0.64  & 1.0  & $3.4 \times 10^{-16}$ & $455^{+20}_{-20}$ \\
42 & (H$\beta$--[O{\sc iii}], [O{\sc ii}], H$\alpha$) & 0.63  & 1.0  & $2.4 \times 10^{-17}$ & $358^{+296}_{-142}$ \\
\hline
\multicolumn{6}{@{}l@{}}{\hbox to 0pt{\parbox{130mm}{\footnotesize
\par\noindent
\footnotemark[$*$] Emission line type. Possible types are put in parentheses.
\par\noindent
\footnotemark[$\dagger$] Photometric redshift.
\par\noindent
\footnotemark[$\ddagger$] FWHM size in the {\it NB816} image.
             The FWHM sizes of stellar objects in the {\it NB816} image are 0$\farcs$9.
\par\noindent
\footnotemark[$\S$] Photometoric error is $\approx 5 \times 10^{-18}$ erg s$^{-1}$ cm$^{-2}$.
\par\noindent
\footnotemark[$\|$] Error shows only the uncertainty of photometry. 
}\hss}}
\end{tabular}
\end{center}
\end{table}

\begin{table}
\begin{center}
\caption{Properties of {\it IA827}-selected strong low-$z$ emitters. \label{tab:iap}}
\begin{tabular}{cccccc}
\hline
\hline
No. & Class\footnotemark[$*$] & $z_{\rm ph}$\footnotemark[$\dagger$] & $FWHM_{\rm obj}$\footnotemark[$\ddagger$] & $F_{\rm line}$\footnotemark[$\S$] & $EW_{\rm obs}$\footnotemark[$\|$] \\
    &       &  & (arcsec) & (erg s$^{-1}$ cm$^{-2}$)  & (\AA) \\
\hline
2 & (H$\beta$--[O{\sc iii}], H$\alpha$, [O{\sc ii}]) & 0.64  & 1.3  & $1.5 \times 10^{-16}$ & $780^{+223}_{-168}$ \\
12 & (H$\alpha$, H$\beta$--[O{\sc iii}]) & 0.25  & 1.8  & $5.5 \times 10^{-17}$ & $802^{+861}_{-390}$ \\
31 & H$\beta$--[O{\sc iii}] & 0.64  & 1.5  & $6.3 \times 10^{-17}$ & $896^{+876}_{-405}$ \\
41 & (H$\beta$--[O{\sc iii}], [O{\sc ii}], H$\alpha$) & 0.63  & 1.3  & $4.9 \times 10^{-16}$ & $779^{+59}_{-55}$ \\
43 & H$\beta$--[O{\sc iii}] & 0.68  & 1.8  & $5.8 \times 10^{-17}$ & $783^{+746}_{-362}$ \\
44 & (H$\beta$--[O{\sc iii}], H$\alpha$) & 0.69  & 1.4  & $9.2 \times 10^{-17}$ & $1086^{+754}_{-398}$ \\
45 & H$\beta$--[O{\sc iii}] & 0.66  & 1.2  & $1.2 \times 10^{-16}$ & $1214^{+665}_{-383}$ \\
46 & H$\alpha$ & 0.26  & 1.6  & $5.2 \times 10^{-17}$ & $800^{+927}_{-403}$ \\
47 & H$\beta$--[O{\sc iii}] & 0.67  & 1.4  & $8.7 \times 10^{-17}$ & $10000^{+\infty}_{-7946}$ \\
48 & H$\beta$--[O{\sc iii}] & 0.67  & 2.0  & $6.7 \times 10^{-17}$ & $1386^{+2402}_{-697}$ \\
49 & H$\beta$--[O{\sc iii}] & 0.69  & 3.5  & $6.1 \times 10^{-17}$ & $750^{+641}_{-331}$ \\
50 & H$\alpha$ & 0.25  & 1.2  & $4.1 \times 10^{-16}$ & $777^{+72}_{-65}$ \\
51 & H$\beta$--[O{\sc iii}] & 0.68  & 1.5  & $7.2 \times 10^{-17}$ & $1087^{+1117}_{-480}$ \\
52 & H$\beta$--[O{\sc iii}] & 0.67  & 1.7  & $6.2 \times 10^{-17}$ & $1042^{+1280}_{-500}$ \\
53 & H$\beta$--[O{\sc iii}] & 0.66  & 1.8  & $1.4 \times 10^{-16}$ & $1781^{+1163}_{-584}$ \\
54 & H$\beta$--[O{\sc iii}] & 0.69  & 2.2  & $5.7 \times 10^{-17}$ & $1112^{+1797}_{-579}$ \\
55 & H$\alpha$ & 0.24  & 2.8  & $6.2 \times 10^{-17}$ & $713^{+568}_{-307}$ \\
\hline
\multicolumn{6}{@{}l@{}}{\hbox to 0pt{\parbox{130mm}{\footnotesize
\par\noindent
\footnotemark[$*$] Emission line type. Possible types are put in parentheses.
\par\noindent
\footnotemark[$\dagger$] Photometric redshift.
\par\noindent
\footnotemark[$\ddagger$] FWHM size in the {\it IA827} image.
             The FWHM sizes of stellar objects in the {\it IA827} image are 1$\farcs$2.
\par\noindent
\footnotemark[$\S$] Photometoric error is $\approx 1.7 \times 10^{-17}$ erg s$^{-1}$ cm$^{-2}$.
\par\noindent
\footnotemark[$\|$] Error shows only the uncertainty of photometry. 
}\hss}}
\end{tabular}
\end{center}
\end{table}

\begin{table}
\begin{center}
\caption{Classification of our emitters.\label{tab:class}}
\begin{tabular}{cccccc}
\hline
\hline
\multicolumn{2}{c}{type} & \multicolumn{4}{c}{ number}                \\
                       &  & {\it NB816} & {\it IA827} & common & total \\
\hline
H$\alpha$              & reliable &  5 &  3 & 1 &  7 \\
                       & possible & 16 &  3 & 2 & 17 \\
                       &  all     & 21 &  6 & 3 & 24 \\
\hline
H$\beta$--[O{\sc iii}] & reliable & 11 & 10 & 1 & 20 \\
                       & possible & 15 &  3 & 2 & 16 \\
                       &  all     & 26 & 13 & 3 & 36 \\
\hline
\ [O{\sc ii}]          & reliable & 10 &  1 & 0 & 11 \\
                       & possible & 11 &  1 & 1 & 11 \\
                       &  all     & 21 &  2 & 1 & 22 \\
\hline
Ly$\alpha$             &          & 20 &  4 & 3 & 21 \\
\hline
\end{tabular}
\end{center}
\end{table}

\begin{table}
\begin{center}
\caption{EW and luminosity properties of H$\alpha$ emitters. \label{tab:ha}}
\begin{tabular}{cccc}
\hline
\hline
No. &  $L_{\rm line}$\footnotemark[$*$] & $EW_0$\footnotemark[$\dagger$] &  $M_B$ \\
    & (erg s$^{-1}$) & (\AA) & \\
\hline
\multicolumn{4}{c}{{\it NB816} -- reliable} \\
\hline
 2 & $2.4 \times 10^{40}$ & $  564^{+   94}_{-   74}$ & -15.5 \\
11 & $9.0 \times 10^{39}$ & $  537^{+  290}_{-  156}$ & -14.8 \\
24 & $6.0 \times 10^{39}$ & $  312^{+  159}_{-   94}$ & -13.6 \\
32 & $5.3 \times 10^{39}$ & $  220^{+   96}_{-   63}$ & -14.1 \\
37 & $2.2 \times 10^{40}$ & $  197^{+   15}_{-   14}$ & -16.1 \\
\hline
\multicolumn{4}{c}{{\it IA827} -- reliable} \\
\hline
 2 & $2.5 \times 10^{40}$ & $  629^{+  178}_{-  135}$ & -14.1 \\
46 & $9.0 \times 10^{39}$ & $  645^{+  748}_{-  324}$ & -16.2 \\
50 & $7.0 \times 10^{40}$ & $  626^{+   57}_{-   52}$ & -14.9 \\
\hline
\multicolumn{4}{c}{{\it NB816} -- possible} \\
\hline
 6 & $3.2 \times 10^{39}$ & $  229^{+  206}_{-   99}$ & -13.9 \\
 7 & $8.9 \times 10^{39}$ & $  250^{+   64}_{-   49}$ & -15.1 \\
 9 & $3.2 \times 10^{39}$ & $  216^{+  188}_{-   93}$ & -14.0 \\
12 & $5.1 \times 10^{39}$ & $  361^{+  274}_{-  132}$ & -13.8 \\
13 & $4.1 \times 10^{39}$ & $  270^{+  206}_{-  104}$ & -14.1 \\
14 & $3.8 \times 10^{39}$ & $  336^{+  376}_{-  147}$ & -14.0 \\
17 & $7.1 \times 10^{39}$ & $  236^{+   76}_{-   55}$ & -14.3 \\
18 & $2.6 \times 10^{39}$ & $  199^{+  209}_{-   96}$ & -13.6 \\
23 & $4.2 \times 10^{39}$ & $  194^{+  103}_{-   65}$ & -14.4 \\
26 & $3.0 \times 10^{39}$ & $  200^{+  168}_{-   87}$ & -13.8 \\
29 & $3.2 \times 10^{39}$ & $  286^{+  352}_{-  135}$ & -13.8 \\
31 & $4.7 \times 10^{39}$ & $  220^{+  111}_{-   70}$ & -14.2 \\
33 & $4.8 \times 10^{39}$ & $  208^{+   97}_{-   63}$ & -14.4 \\
35 & $3.1 \times 10^{39}$ & $  240^{+  243}_{-  109}$ & -14.1 \\
38 & $1.4 \times 10^{40}$ & $  209^{+   29}_{-   25}$ & -15.7 \\
42 & $4.1 \times 10^{39}$ & $  288^{+  238}_{-  115}$ & -13.5 \\
\hline
\multicolumn{4}{c}{{\it IA827} -- possible} \\
\hline
12 & $9.4 \times 10^{39}$ & $  646^{+  694}_{-  314}$ & -13.9 \\
31 & $1.1 \times 10^{40}$ & $  722^{+  705}_{-  326}$ & -13.8 \\
44 & $1.6 \times 10^{40}$ & $  875^{+  608}_{-  320}$ & -14.1 \\
\hline
\multicolumn{4}{@{}l@{}}{\hbox to 0pt{\parbox{130mm}{\footnotesize
\par\noindent
\footnotemark[$*$] Photometoric error is
  $\approx 8 \times 10^{38}$ erg s$^{-1}$ for {\it NB816} and
  $\approx 3.0 \times 10^{39}$ erg s$^{-1}$ for {\it IA827}.
\par\noindent
\footnotemark[$\dagger$] The error shows only the uncertainty of photometry. 
}\hss}}
\end{tabular}
\end{center}
\end{table}

\begin{table}
\begin{center}
\caption{EW and luminosity properties of H$\beta$ -- [O{\sc iii}] emitters. \label{tab:hb}}
\renewcommand\arraystretch{0.75}
\begin{tabular}{cccc}
\hline
\hline
No. &  $L_{\rm line}$\footnotemark[$*$] & $EW_0$\footnotemark[$\dagger$] &  $M_B$ \\
    & (erg s$^{-1}$) & (\AA) & \\
\hline
\multicolumn{4}{c}{{\it NB816} -- reliable} \\
\hline
 4 & $6.7 \times 10^{41}$ & $  303^{+   12}_{-   12}$ & -19.4 \\
10 & $7.4 \times 10^{41}$ & $  223^{+    6}_{-    6}$ & -19.8 \\
21 & $1.1 \times 10^{41}$ & $  290^{+   88}_{-   61}$ & -17.3 \\
22 & $1.2 \times 10^{41}$ & $  174^{+   30}_{-   25}$ & -17.9 \\
25 & $1.4 \times 10^{41}$ & $  256^{+   49}_{-   39}$ & -17.7 \\
28 & $6.4 \times 10^{41}$ & $  163^{+    4}_{-    4}$ & -19.8 \\
30 & $3.2 \times 10^{41}$ & $  147^{+    8}_{-    7}$ & -19.1 \\
34 & $2.2 \times 10^{41}$ & $  448^{+   87}_{-   66}$ & -17.7 \\
36 & $2.9 \times 10^{41}$ & $  281^{+   26}_{-   23}$ & -18.4 \\
39 & $7.8 \times 10^{41}$ & $  293^{+   10}_{-    9}$ & -19.4 \\
41 & $5.9 \times 10^{41}$ & $  275^{+   11}_{-   12}$ & -19.2 \\
\hline
\multicolumn{4}{c}{{\it IA827} -- reliable} \\
\hline
41 & $8.5 \times 10^{41}$ & $  472^{+   35}_{-   33}$ & -17.0 \\
43 & $1.0 \times 10^{41}$ & $  474^{+  452}_{-  219}$ & -16.4 \\
45 & $2.0 \times 10^{41}$ & $  735^{+  403}_{-  232}$ & -17.0 \\
47 & $1.5 \times 10^{41}$ & $ 6060^{+\infty}_{-4815}$ & -16.9 \\
48 & $1.2 \times 10^{41}$ & $  840^{+ 1455}_{-  422}$ & -16.7 \\
49 & $1.0 \times 10^{41}$ & $  454^{+  387}_{-  200}$ & -17.3 \\
51 & $1.2 \times 10^{41}$ & $  658^{+  676}_{-  290}$ & -16.8 \\
52 & $1.1 \times 10^{41}$ & $  631^{+  775}_{-  303}$ & -18.0 \\
53 & $2.5 \times 10^{41}$ & $ 1079^{+  704}_{-  353}$ & -16.7 \\
54 & $9.8 \times 10^{40}$ & $  673^{+ 1088}_{-  350}$ & -17.0 \\
\hline
\multicolumn{4}{c}{{\it NB816} -- possible} \\
\hline
 6 & $3.2 \times 10^{40}$ & $  172^{+  155}_{-   75}$ & -16.4 \\
 7 & $9.0 \times 10^{40}$ & $  188^{+   48}_{-   36}$ & -17.6 \\
 9 & $3.2 \times 10^{40}$ & $  163^{+  141}_{-   70}$ & -16.4 \\
12 & $5.1 \times 10^{40}$ & $  271^{+  206}_{-   99}$ & -16.5 \\
13 & $4.1 \times 10^{40}$ & $  203^{+  155}_{-   78}$ & -16.5 \\
14 & $3.8 \times 10^{40}$ & $  252^{+  282}_{-  110}$ & -16.3 \\
17 & $7.1 \times 10^{40}$ & $  177^{+   57}_{-   41}$ & -17.2 \\
18 & $2.6 \times 10^{40}$ & $  149^{+  157}_{-   72}$ & -16.3 \\
23 & $4.2 \times 10^{40}$ & $  146^{+   78}_{-   49}$ & -16.9 \\
26 & $3.1 \times 10^{40}$ & $  150^{+  126}_{-   65}$ & -16.4 \\
29 & $3.2 \times 10^{40}$ & $  215^{+  264}_{-  101}$ & -16.3 \\
31 & $4.7 \times 10^{40}$ & $  166^{+   83}_{-   52}$ & -17.0 \\
35 & $3.1 \times 10^{40}$ & $  180^{+  183}_{-   82}$ & -16.3 \\
38 & $1.4 \times 10^{41}$ & $  157^{+   22}_{-   19}$ & -18.2 \\
42 & $4.1 \times 10^{40}$ & $  216^{+  179}_{-   86}$ & -16.4 \\
\hline
\multicolumn{4}{c}{{\it IA827} -- possible} \\
\hline
12 & $9.5 \times 10^{40}$ & $  486^{+  521}_{-  236}$ & -17.1 \\
31 & $1.1 \times 10^{41}$ & $  543^{+  530}_{-  245}$ & -17.2 \\
44 & $1.6 \times 10^{41}$ & $  658^{+  457}_{-  241}$ & -16.7 \\
\hline
\multicolumn{4}{@{}l@{}}{\hbox to 0pt{\parbox{130mm}{\footnotesize
\par\noindent
\footnotemark[$*$] Photometoric error is
  $\approx 8 \times 10^{39}$ erg s$^{-1}$ for {\it NB816} and
  $\approx 3.0 \times 10^{40}$ erg s$^{-1}$ for {\it IA827}.
\par\noindent
\footnotemark[$\dagger$] The error shows only the uncertainty of photometry. 
}\hss}}
\end{tabular}
\end{center}
\end{table}

\begin{table}
\begin{center}
\caption{EW and luminosity properties of [O{\sc ii}] emitters. \label{tab:oii}}
\begin{tabular}{cccc}
\hline
\hline
No. &  $L_{\rm line}$\footnotemark[$*$] & $EW_0$\footnotemark[$\dagger$] &  $M_{410}$ \\
    & (erg s$^{-1}$) & (\AA) & \\
\hline
\multicolumn{4}{c}{{\it NB816} -- reliable} \\
\hline
 1 & $8.7 \times 10^{41}$ & $  117^{+   11}_{-   10}$ & -20.1 \\
 3 & $2.6 \times 10^{41}$ & $  124^{+   52}_{-   35}$ & -18.6 \\
 5 & $1.9 \times 10^{41}$ & $  112^{+   62}_{-   38}$ & -18.6 \\
 8 & $1.9 \times 10^{42}$ & $  164^{+    8}_{-    8}$ & -20.3 \\
15 & $2.3 \times 10^{42}$ & $  115^{+    4}_{-    4}$ & -20.7 \\
16 & $3.0 \times 10^{41}$ & $  117^{+   39}_{-   28}$ & -18.9 \\
19 & $2.7 \times 10^{41}$ & $  193^{+  114}_{-   62}$ & -18.4 \\
20 & $3.7 \times 10^{41}$ & $  126^{+   34}_{-   26}$ & -19.2 \\
27 & $3.1 \times 10^{41}$ & $  119^{+   37}_{-   27}$ & -18.9 \\
40 & $4.8 \times 10^{41}$ & $  117^{+   22}_{-   18}$ & -19.3 \\
\hline
\multicolumn{4}{c}{{\it IA827} -- reliable} \\
\hline
55 & $4.9 \times 10^{41}$ & $  325^{+  259}_{-  140}$ & -20.3 \\
\hline
\multicolumn{4}{c}{{\it NB816} -- possible} \\
\hline
 6 & $1.5 \times 10^{41}$ & $  129^{+  117}_{-   56}$ & -18.1 \\
 9 & $1.5 \times 10^{41}$ & $  122^{+  106}_{-   53}$ & -18.3 \\
12 & $2.4 \times 10^{41}$ & $  204^{+  155}_{-   74}$ & -17.8 \\
13 & $1.9 \times 10^{41}$ & $  152^{+  116}_{-   59}$ & -18.1 \\
14 & $1.8 \times 10^{41}$ & $  190^{+  213}_{-   83}$ & -18.2 \\
18 & $1.2 \times 10^{41}$ & $  112^{+  118}_{-   54}$ & -17.9 \\
26 & $1.4 \times 10^{41}$ & $  113^{+   95}_{-   49}$ & -18.2 \\
29 & $1.5 \times 10^{41}$ & $  162^{+  199}_{-   76}$ & -17.6 \\
33 & $2.2 \times 10^{41}$ & $  117^{+   55}_{-   36}$ & -18.4 \\
35 & $1.4 \times 10^{41}$ & $  136^{+  138}_{-   61}$ & -18.2 \\
42 & $1.9 \times 10^{41}$ & $  163^{+  134}_{-   65}$ & -17.9 \\
\hline
\multicolumn{4}{c}{{\it IA827} -- possible} \\
\hline
12 & $4.4 \times 10^{41}$ & $  366^{+  393}_{-  177}$ & -18.0 \\
\hline
\multicolumn{4}{@{}l@{}}{\hbox to 0pt{\parbox{130mm}{\footnotesize
\par\noindent
\footnotemark[$*$] Photometoric error is
  $\approx 4 \times 10^{40}$ erg s$^{-1}$ for {\it NB816} and
  $\approx 1.4 \times 10^{41}$ erg s$^{-1}$ for {\it IA827}.
\par\noindent
\footnotemark[$\dagger$] The error shows only the uncertainty of photometry. 
}\hss}}
\end{tabular}
\end{center}
\end{table}

\begin{table}
\begin{center}
\caption{EW and luminosity properties of {\it NB816}-selected strong low-$z$ emitters. \label{tab:survey}}
\begin{tabular}{cccccccc} 
\hline
\hline
catalog     &  redshift range & volume          & $EW_{\rm 0,lim}$ & $M_{\rm lim}$\footnotemark[$*$] & Number\footnotemark[$\dagger$] & Number density\footnotemark[$\dagger$] \\
            &                 & (Mpc$^3$) & \AA          &        &       & (Mpc$^{-3}$) \\
\hline
\multicolumn{7}{c}{H$\alpha$} \\
\hline
{\it NB816} &  0.23 -- 0.25  & $4.0\times10^3$ & $\approx 190$ & $M_B \approx -13.5$ & 5--21 & 1.2--$5.2 \times 10^{-3}$ \\
{\it IA827} &  0.23 -- 0.29  & $1.1\times10^4$ & $\approx 550$ & $M_B \approx -13.5$ & 3--6  & 2.7--$5.4 \times 10^{-4}$ \\
 SDSS DR3   & 0.003 -- 0.006 & $6.3\times10^3$ &          190  & $M_B = -13.5$ &   62  &      $1.0 \times 10^{-2}$ \\
 SDSS DR3   & 0.003 -- 0.006 & $6.3\times10^3$ &          550  & $M_B = -13.5$ &   18  &      $3.2 \times 10^{-3}$ \\
\hline
\multicolumn{7}{c}{H$\beta$--[O{\sc iii}]} \\
\hline
{\it NB816} &  0.62 -- 0.69  & $6.6\times10^4$ & $\approx 140$ & $M_B \approx -16.3$ & 11--26 & 1.7--$4.0 \times 10^{-4}$ \\
{\it IA827} &  0.62 -- 0.73  & $1.3\times10^5$ & $\approx 410$ & $M_B \approx -16.3$ & 10--13 & 7.6--$10  \times 10^{-5}$ \\
SDSS DR3    & 0.008 -- 0.015 & $9.5\times10^4$ &          140  & $M_B = -16.3$ &    55  &      $5.5 \times 10^{-4}$ \\
SDSS DR3    & 0.008 -- 0.015 & $9.5\times10^4$ &          410  & $M_B = -16.3$ &    13  &      $1.3 \times 10^{-4}$ \\
\hline
\multicolumn{7}{c}{[O{\sc ii}]} \\
\hline
{\it NB816} &  1.17 -- 1.20  & $6.0\times10^5$ & $\approx 110$ & $M_{410}\approx-17.6$ & 10--21 & 1.7--$3.5 \times 10^{-4}$ \\
{\it IA827} &  1.17 -- 1.26  & $1.7\times10^6$ & $\approx 310$ & $M_{410}\approx-17.6$ &  1--2  & 5.9--$12 \times 10^{-7}$ \\
SDSS DR3    &  0.03 -- 0.06  & $6.0\times10^6$ &          110  & $M_{410}=-17.6$ &   178  & $3.0 \times 10^{-5}$ \\
SDSS DR3    &  0.03 -- 0.06  & $6.0\times10^6$ &          310  & $M_{410}=-17.6$ &    22  & $3.3 \times 10^{-6}$ \\
\hline
\multicolumn{7}{@{}l@{}}{\hbox to 0pt{\parbox{130mm}{\footnotesize
\par\noindent
\footnotemark[$*$] Limiting absolute magnitude.
\par\noindent
\footnotemark[$\dagger$] The cases for only reliable sample and all (reliable and possible) sample are
                         shown as the minimum and maximum values, respectively.
}\hss}}
\end{tabular}
\end{center}
\end{table}

\begin{figure}
\FigureFile(160mm,250mm){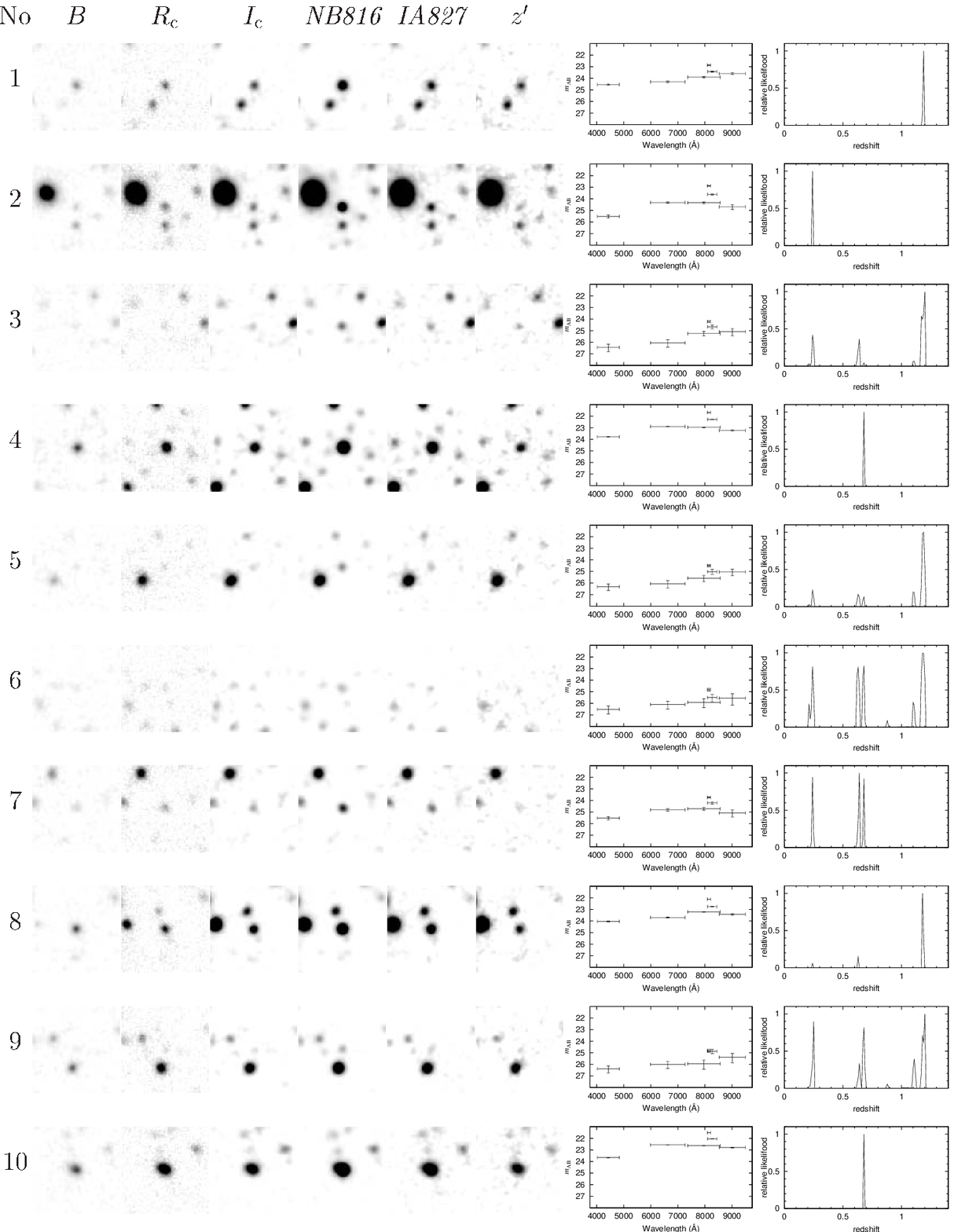} 
\end{figure}
\begin{figure}
\FigureFile(160mm,250mm){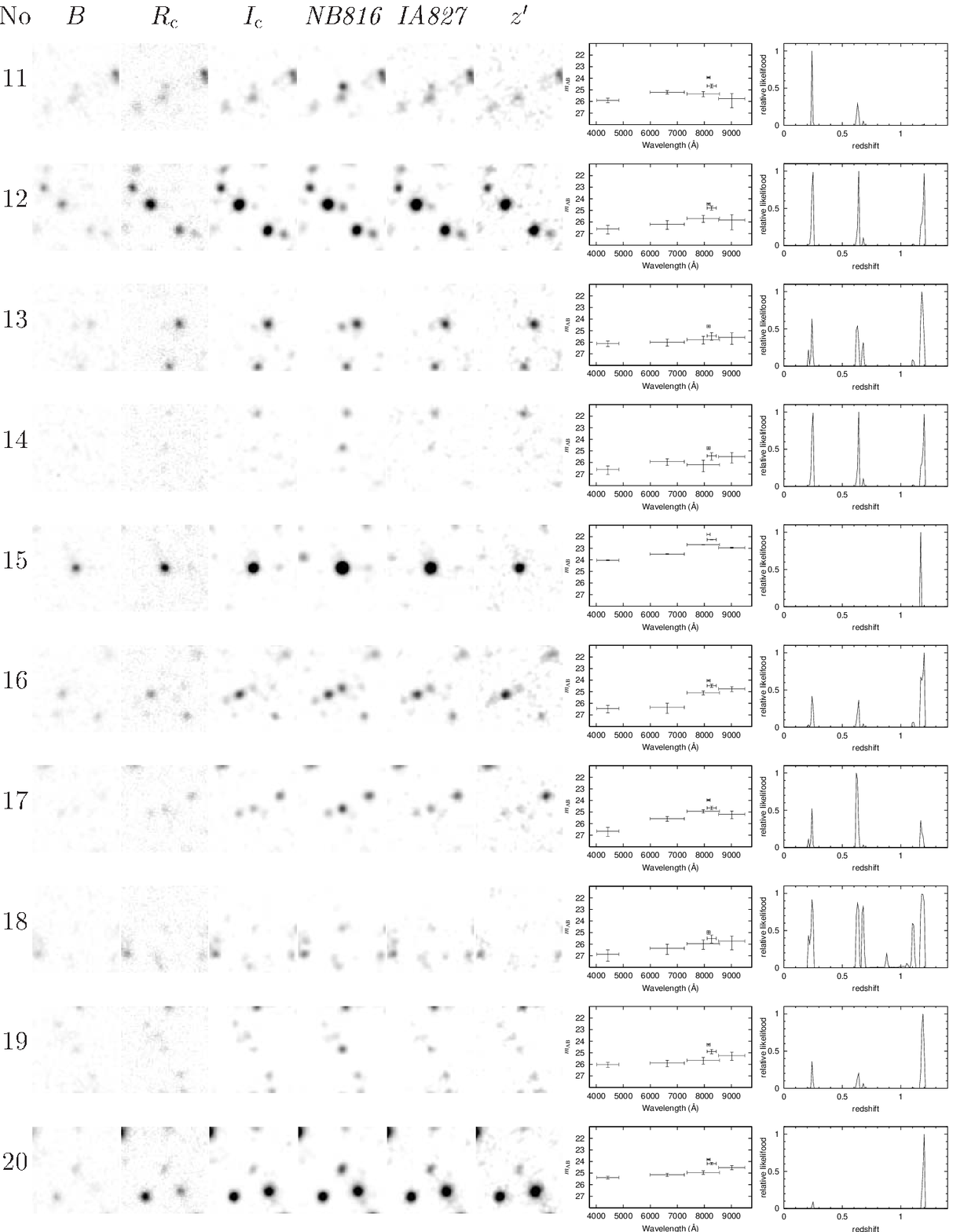} 
\end{figure}
\begin{figure}
\FigureFile(160mm,250mm){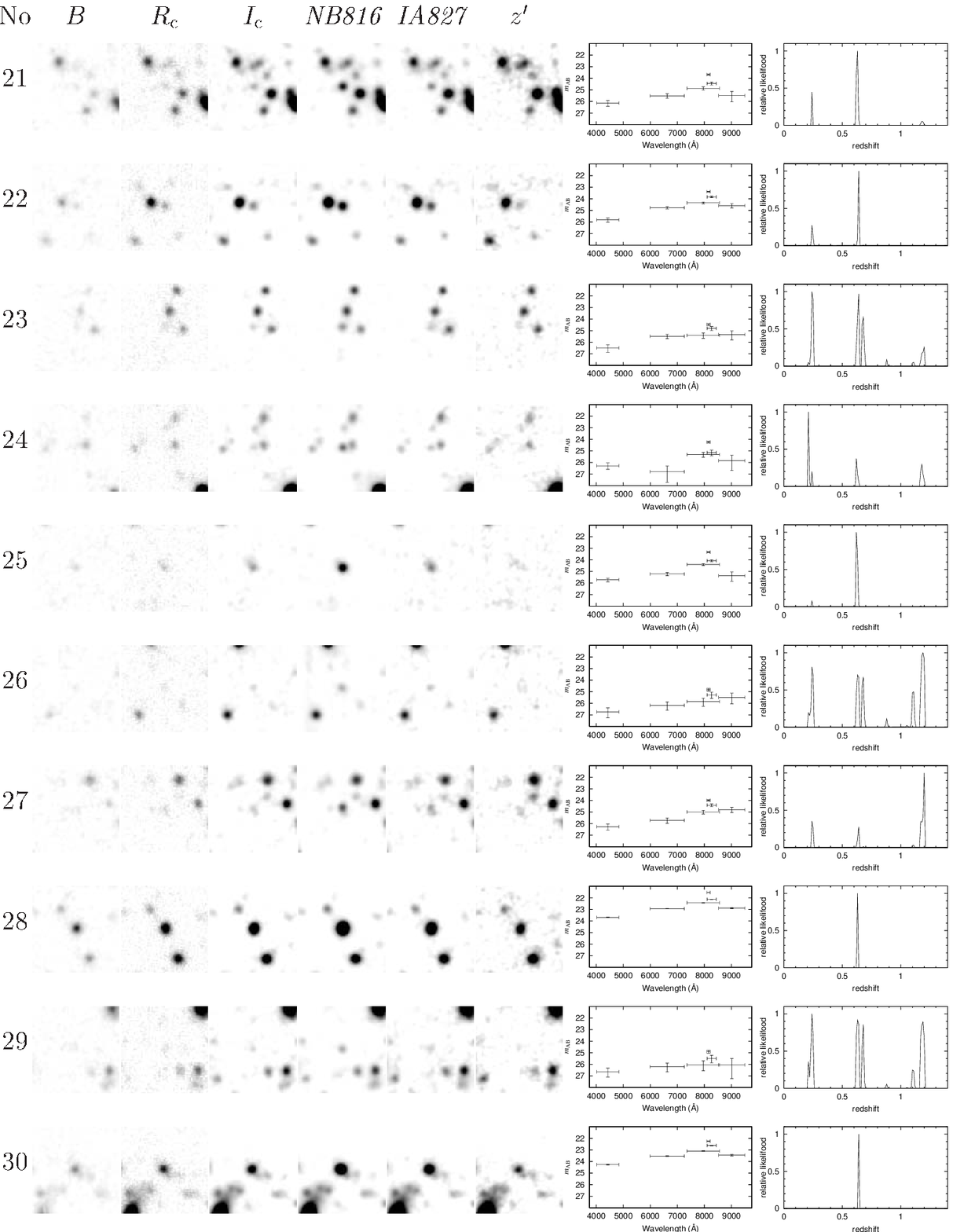} 
\end{figure}
\begin{figure}
\FigureFile(160mm,250mm){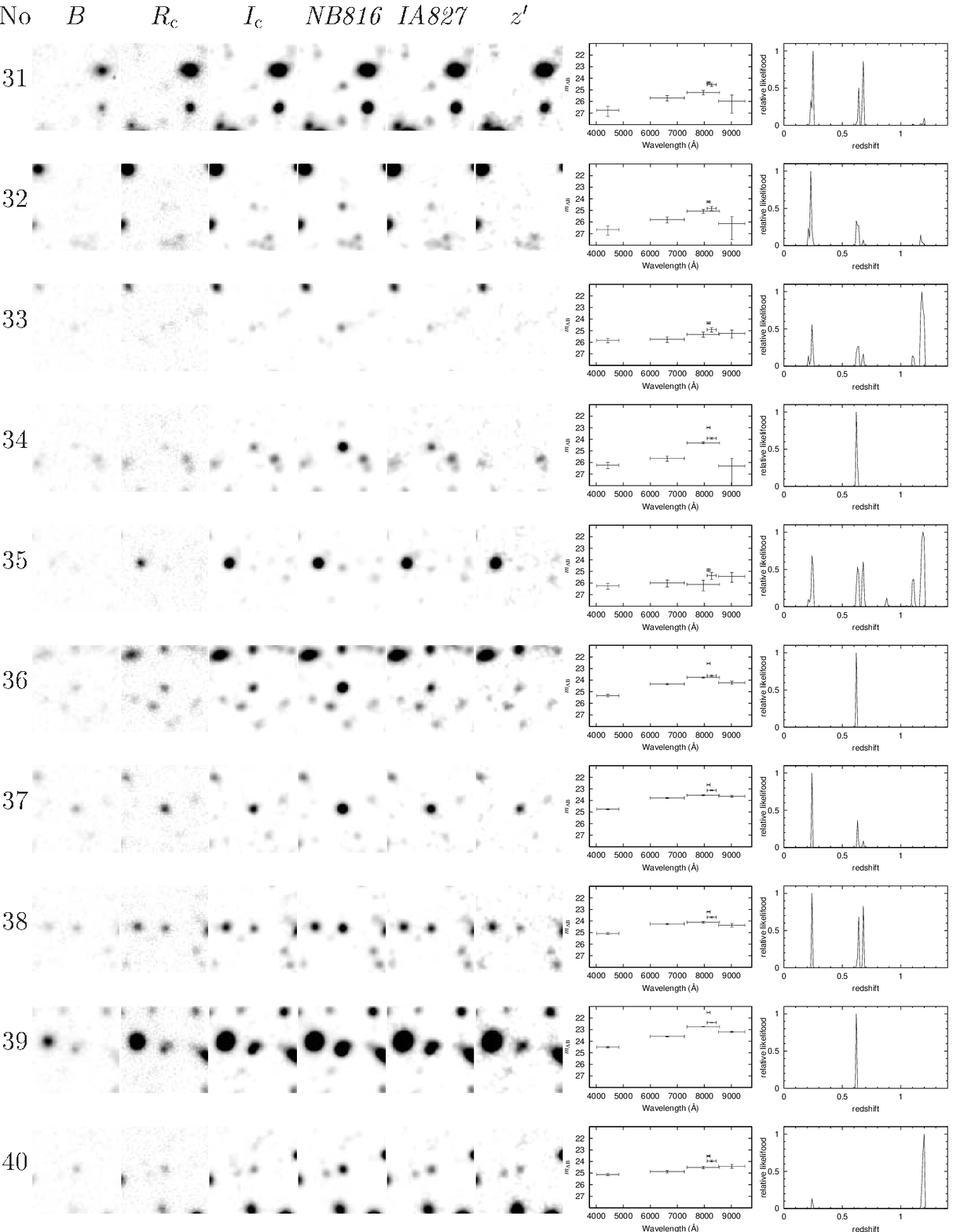} 
\end{figure}

\begin{figure}
\FigureFile(160mm,250mm){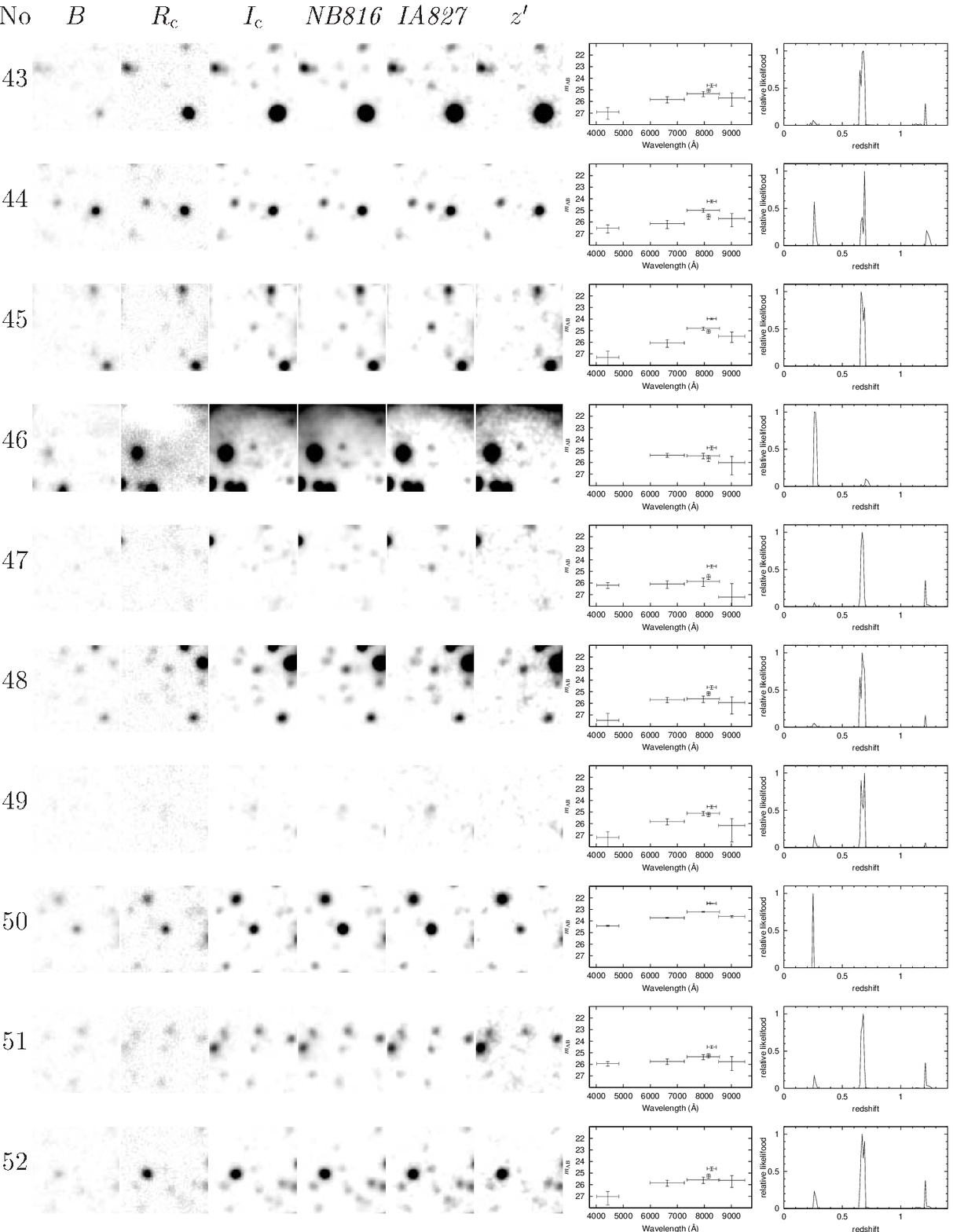} 
\end{figure}

\end{document}